\pdfoutput=1
\PassOptionsToPackage{unicode}{hyperref}
\PassOptionsToPackage{hyphens}{url}
\documentclass[
  a4paper,
  11pt]{article}
\usepackage{xcolor}
\usepackage{amsmath,amssymb}
\usepackage{iftex}
\ifPDFTeX
  \usepackage[T1]{fontenc}
  \usepackage[utf8]{inputenc}
  \usepackage{textcomp} % provide euro and other symbols
\else % if luatex or xetex
  \usepackage{unicode-math} % this also loads fontspec
  \defaultfontfeatures{Scale=MatchLowercase}
  \defaultfontfeatures[\rmfamily]{Ligatures=TeX,Scale=1}
\fi
\usepackage{lmodern}
\ifPDFTeX\else
\fi
\IfFileExists{upquote.sty}{\usepackage{upquote}}{}
\IfFileExists{microtype.sty}{% use microtype if available
  \usepackage[]{microtype}
  \UseMicrotypeSet[protrusion]{basicmath} % disable protrusion for tt fonts
}{}
\makeatletter
\@ifundefined{KOMAClassName}{% if non-KOMA class
  \IfFileExists{parskip.sty}{%
    \usepackage{parskip}
  }{% else
    \setlength{\parindent}{0pt}
    \setlength{\parskip}{6pt plus 2pt minus 1pt}}
}{% if KOMA class
  \KOMAoptions{parskip=half}}
\makeatother
\usepackage{longtable,booktabs,array}
\usepackage{calc} % for calculating minipage widths
\usepackage{etoolbox}
\makeatletter
\patchcmd\longtable{\par}{\if@noskipsec\mbox{}\fi\par}{}{}
\makeatother
\IfFileExists{footnotehyper.sty}{\usepackage{footnotehyper}}{\usepackage{footnote}}
\makesavenoteenv{longtable}
\usepackage{graphicx}
\makeatletter
\newsavebox\pandoc@box
\newcommand*\pandocbounded[1]{% scales image to fit in text height/width
  \sbox\pandoc@box{#1}%
  \Gscale@div\@tempa{\textheight}{\dimexpr\ht\pandoc@box+\dp\pandoc@box\relax}%
  \Gscale@div\@tempb{\linewidth}{\wd\pandoc@box}%
  \ifdim\@tempb\p@<\@tempa\p@\let\@tempa\@tempb\fi% select the smaller of both
  \ifdim\@tempa\p@<\p@\scalebox{\@tempa}{\usebox\pandoc@box}%
  \else\usebox{\pandoc@box}%
  \fi%
}
\def\fps@figure{htbp}
\makeatother
\providecommand{\tightlist}{%
  \setlength{\itemsep}{0pt}\setlength{\parskip}{0pt}}
\usepackage[margin=1in]{geometry}

\usepackage{fancyhdr}
\AtBeginDocument{%
  \hypersetup{colorlinks=true,linkcolor=blue!70!black,citecolor=blue!70!black,urlcolor=blue!70!black}%
}

\usepackage{booktabs}

\usepackage{bookmark}
\IfFileExists{xurl.sty}{\usepackage{xurl}}{} % add URL line breaks if available
\hypersetup{
  hidelinks,
  pdfcreator={LaTeX via pandoc}}

\author{}
\date{}

\begin{document}

\section{RLM-on-KG: Heuristics First, LLMs When
Needed}\label{rlm-on-kg-heuristics-first-llms-when-needed}

\subsection{Adaptive Retrieval Control over Mention Graphs for Scattered
Evidence}\label{adaptive-retrieval-control-over-mention-graphs-for-scattered-evidence}

\textbf{Andrea Volpini} (andrea@wordlift.io), \textbf{Elie Raad}
(elie@wordlift.io) WordLift

\begin{center}\rule{0.5\linewidth}{0.5pt}\end{center}

\subsection{Abstract}\label{abstract}

When does an LLM controller outperform rule-based traversal for
knowledge graph exploration? We study this question through RLM-on-KG, a
retrieval system that treats an LLM as an autonomous navigator over an
RDF-encoded mention graph for grounded question answering. Unlike
GraphRAG pipelines that rely on offline LLM indexing, RLM-on-KG performs
entity-first, multi-hop exploration at query time using deterministic
graph construction and a fixed tool set.

Our central finding is a \textbf{conditional advantage}: the LLM
controller's value depends on two factors --- the scatter of required
evidence and the model's tool-calling sophistication. \textbf{The
paper's core claim is LLM control versus heuristic traversal}, not a
generic win over GraphRAG. On GraphRAG-Bench Novel (519 questions), the
Gemini 2.0 Flash controller achieves +2.47 pp F1 over a rule-based
heuristic baseline (p \textless{} 0.0001, 95\% CI {[}+1.41, +3.56{]});
against the GraphRAG-local variant, the same controller achieves only
+0.16 pp (p=0.36, not significant). With a more capable controller ---
Claude Haiku 4.5 on the same 519 questions --- the heuristic advantage
grows to \textbf{+4.37 pp} (p \textless{} 0.001) and extends to a
\textbf{+2.42 pp statistically significant win over the controlled
GraphRAG-local variant} (p \textless{} 0.001; 170 wins, 259 ties, 90
losses): the first significant improvement over entity-expanded
retrieval, contingent on strong tool-calling capability. The gain is
largest when gold evidence is scattered across 6--10 chunks (+3.21 pp,
65\% win rate) and smallest for concentrated evidence (+1.85 pp). Across
model families (different experiment scales --- see Section 7 for the
matched-scale comparison): Claude (+4.37 pp, n=519) \textgreater{}
Gemini (+0.84 pp, n=100) \textgreater{} Gemma 4 E2B ($-$0.78 pp, n=100,
83\% behavioral tie rate) --- establishing a clear distillation target
for future work. Cross-scale validation on MuSiQue confirms the
LLM-over-heuristic advantage transfers, with expected attenuation on
smaller per-question graphs.

The core architectural insight is the separation of \textbf{candidate
discovery} (LLM-driven graph traversal) from \textbf{ranking} (pure
vector re-ranking): the LLM adds value through exploration breadth, not
judgment.

Beyond retrieval, exploration traces serve as a proposed
\textbf{stress-test harness for structured data quality}, producing
diagnostics for the underlying graph substrate (coverage, connectivity,
provenance, and queryability).

\begin{center}\rule{0.5\linewidth}{0.5pt}\end{center}

\subsection{1. Introduction}\label{introduction}

LLM-guided graph traversal is costly and unnecessary for many retrieval
problems. A pure vector search handles single-hop factoid questions. A
one-hop entity expansion (GraphRAG-local) handles moderate multi-hop
questions. So when, exactly, does deploying an LLM as an active graph
navigator become worth the additional latency and token cost?

We study this question by placing four retrieval strategies on a
continuum of controller complexity:

{\def\LTcaptype{none} % do not increment counter
\begin{longtable}[]{@{}
  >{\raggedright\arraybackslash}p{(\linewidth - 4\tabcolsep) * \real{0.3929}}
  >{\raggedright\arraybackslash}p{(\linewidth - 4\tabcolsep) * \real{0.3929}}
  >{\raggedright\arraybackslash}p{(\linewidth - 4\tabcolsep) * \real{0.2143}}@{}}
\toprule\noalign{}
\begin{minipage}[b]{\linewidth}\raggedright
Controller
\end{minipage} & \begin{minipage}[b]{\linewidth}\raggedright
Mechanism
\end{minipage} & \begin{minipage}[b]{\linewidth}\raggedright
Cost
\end{minipage} \\
\midrule\noalign{}
\endhead
\bottomrule\noalign{}
\endlastfoot
Vector-only & top-k embedding similarity & \textasciitilde2K tokens,
\textless1s \\
GraphRAG-local & 1-hop entity expansion + vector re-rank &
\textasciitilde2K tokens, \textless1s \\
Heuristic RLM & BFS multi-hop traversal, fixed rules & \textasciitilde5K
tokens, \textasciitilde5s \\
\textbf{LLM RLM} & Adaptive multi-hop, LLM-driven tool selection &
\textasciitilde50K tokens, 2--5 min \\
\end{longtable}
}

All four share the same deterministically-built mention graph, the same
embedding index, and the same evaluation pipeline. The only variable is
the controller.

Retrieval-augmented generation (RAG) has become the standard approach
for grounding LLM responses in external knowledge (Lewis et al., 2020).
The core recipe is straightforward: embed a query, retrieve the top-k
most similar chunks from a vector index, and pass them as context to a
generator. This works well for single-hop factoid questions, but breaks
down when answers require synthesizing information scattered across
multiple document sections, entities, and relationships.

GraphRAG (Edge et al., 2024) addresses this by layering entity-based
expansion on top of vector retrieval. In its full pipeline, GraphRAG
uses an LLM to extract entities, builds a community graph via the Leiden
algorithm, and generates community summaries at indexing time. At query
time, these precomputed structures enable multi-hop retrieval by
expanding from seed chunks to related entities and their neighborhoods.
This improves recall on complex questions, but at significant offline
cost: LLM-based entity extraction, community detection, and summary
generation must be performed before the first query can be answered. For
evolving corpora, this pipeline must be re-run whenever the underlying
documents change.

Recursive Language Models (RLMs) offer an alternative paradigm.
Introduced by Zhang, Kraska, and Khattab (2025), RLMs treat the LLM
itself as an environment explorer rather than a passive consumer of
retrieved context. The LLM proposes actions (tool calls), observes
results, and decides what to explore next, iterating until it has
gathered sufficient evidence. The original RLM work demonstrated this on
Python REPLs and web browsing; we adapt it to mention-graph traversal.
This setting is motivated by RDF-based knowledge graphs, such as the
WordLift KG, where entity-centric identifiers, provenance links to
source text, and structured publishing workflows are already available
and can be reused as a retrieval substrate.

\textbf{Our contribution.} We present five findings:

\begin{enumerate}
\def\labelenumi{\arabic{enumi}.}
\item
  \textbf{LLM control vs.~heuristic traversal.} On the same mention
  graph and tool set, LLM-driven exploration statistically significantly
  outperforms rule-based heuristic traversal: +2.47 pp F1 with Gemini (p
  \textless{} 0.0001, n=519), +4.37 pp with Claude (p \textless{} 0.001,
  n=519).
\item
  \textbf{Conditional advantage.} The gain is not uniform but
  conditional on evidence scatter: +3.21 pp when gold evidence spans
  6--10 chunks, +1.85 pp when concentrated. The LLM adds value precisely
  when graph structure demands adaptive navigation.
\item
  \textbf{Cross-model scaling.} The advantage scales with tool-calling
  sophistication: Claude (+4.37 pp, n=519) \textgreater{} Gemini (+0.84
  pp, n=100) \textgreater{} Gemma 4 E2B ($-$0.78 pp, n=100). These deltas
  span different experiment scales; at matched n=100, only Claude
  achieves a statistically significant advantage (see Section 7 for the
  matched-scale comparison). At full scale (n=519), Claude is the first
  controller to statistically significantly outperform the
  GraphRAG-local variant (+2.42 pp, p \textless{} 0.001).
\item
  \textbf{Discovery vs.~ranking separation.} The LLM adds value through
  candidate discovery, not ranking judgment. Pure vector re-ranking of
  LLM-discovered candidates outperforms all composite scoring strategies
  (Section 5.2).
\item
  \textbf{Traces as supervision.} Exploration traces from strong
  controllers \emph{can} serve as a proposed KG diagnostics framework
  and training signal for distilling navigation policies into smaller
  models (both proposed; empirical validation is future work).
\end{enumerate}

\textbf{The retrieval loop.} At its core, RLM-on-KG implements a cyclic
retrieval process (Figure 1): the LLM seeds entities from the question,
expands neighborhoods via mention links, reads and verifies chunks, then
either collects evidence and cites, or refines sub-queries and loops
back.

\begin{figure}
\centering
\pandocbounded{\includegraphics[keepaspectratio,alt={The retrieval loop. Each stage is an optimization target: seed entity resolution depends on entity identifiers, expansion depends on mention links, verification depends on provenance, and citation depends on stable chunk anchors.}]{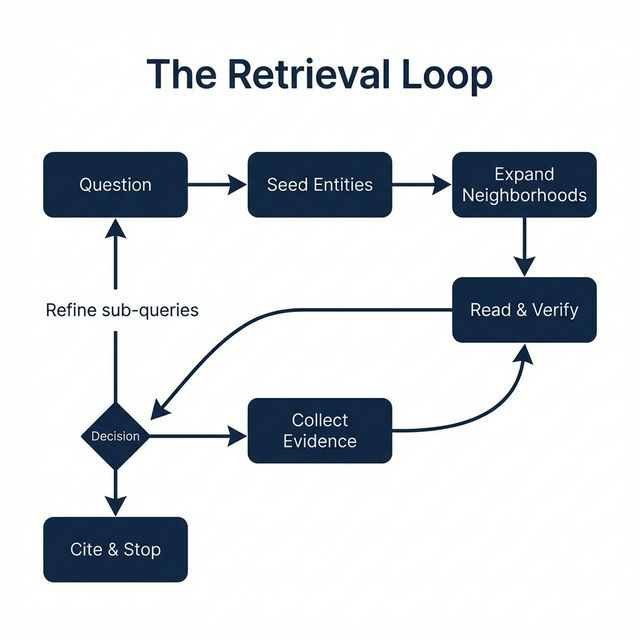}}
\caption{The retrieval loop. Each stage is an optimization target: seed
entity resolution depends on entity identifiers, expansion depends on
mention links, verification depends on provenance, and citation depends
on stable chunk anchors.}
\end{figure}

\textbf{Terminology.} Throughout this paper, ``mention-graph traversal''
refers to traversal over the co-mention graph stored in RDF. We use
``WordLift Knowledge Graph'' only for the platform layer and identifiers
(Section 3.6). Where we use the shorthand ``KG traversal,'' we mean
traversal over this mention graph.

\begin{center}\rule{0.5\linewidth}{0.5pt}\end{center}

\subsection{2. Related Work}\label{related-work}

\textbf{Retrieval-augmented generation.} Lewis et al.~(2020) introduced
RAG as a way to ground generative models in retrieved passages.
Subsequent work has improved retrieval quality through better embeddings
(Izacard and Grave, 2021), query rewriting (Ma et al., 2023), and
multi-step retrieval (Trivedi et al., 2023). RLM-on-KG builds on this
tradition but replaces fixed retrieval pipelines with LLM-driven
exploration.

\textbf{GraphRAG.} Edge et al.~(2024) proposed a graph-based approach to
RAG that constructs an entity graph from documents, detects communities,
and generates summaries at each level. The system operates in two modes:
\emph{local} retrieval (entity expansion from seed chunks) and
\emph{global} summarization (traversing community summaries). Our
GraphRAG-local variant implements the local recipe without community
detection or community summaries; see Section 3.5 for which components
are intentionally excluded and why. The full GraphRAG pipeline requires
substantial offline LLM processing; our approach avoids this entirely.

\textbf{Contemporary graph retrievers.} Several recent systems address
graph-structured retrieval with complementary approaches. HippoRAG
(Gutierrez et al., NeurIPS 2024) uses an LLM to build a
hippocampus-inspired knowledge graph offline and performs Personalized
PageRank at query time. IRCoT (Trivedi et al., ACL 2023) interleaves
retrieval and chain-of-thought steps without an explicit graph
structure. LightRAG (Guo et al., 2025) constructs a lightweight
entity-relation graph with incremental updates and performs dual-mode
retrieval (local entity expansion and global keyword search). Our
comparison design intentionally holds graph construction, embedding
model, chunk budget, and generator fixed to isolate the retrieval logic
itself; integrating these external systems as baselines would conflate
retrieval policy differences with differences in graph construction,
NER, and embedding, making attributions impossible. Our heuristic RLM
baseline (Section 4.4) plays a role analogous to a ReAct-style agent
baseline: same tools and graph, but rule-based tool selection, isolating
the contribution of LLM-driven control.

\textbf{Generative Knowledge Extraction and Graph Reasoning.} Buehler
(2024) transformed scientific literature into an ontological knowledge
graph and used structural analysis (node degrees, isomorphisms, path
sampling) to synthesize novel material designs and interdisciplinary
relationships. While they perform structural analysis offline to
discover insights, RLM-on-KG adapts these principles to query-time
retrieval, using an LLM to actively navigate structural relationships
and gather scattered evidence.

\textbf{Recursive Language Models.} Zhang, Kraska, and Khattab (2025)
introduced the RLM framework, where an LLM recursively applies tools to
explore an environment. Their work demonstrated the approach on Python
REPLs and web browsing tasks. We adapt the core RLM idea to knowledge
graph traversal, replacing the REPL with KG tools and adding
domain-specific scoring and selection mechanisms.

\textbf{Adaptive computation and looped models.} Ouro (Zhu et al., 2025)
demonstrates that \emph{adaptive iteration depth} improves reasoning: a
looped transformer applies the same block recurrently, with an
entropy-regularized objective that allocates more iterations to complex
inputs and fewer to simple ones. RLM-on-KG performs the analogous
optimization over external structure rather than latent space: it
applies the same LLM-driven tool-calling loop over a mention graph, with
evidence scatter determining how many hops are needed. Simple questions
saturate after 1--2 hops; complex scattered evidence requires 8--12.
This connection positions adaptive retrieval depth as the external
analog of adaptive computational depth.

\textbf{Graph reasoning in context.} The GraphWalks benchmark (OpenAI,
2025) tests whether LLMs can perform structured graph traversal (BFS,
parent retrieval) when a graph is placed in-context. Performance
degrades sharply as graph size grows, confirming that in-context
multi-hop reasoning over graph structure does not scale. RLM-on-KG
sidesteps this limitation by externalizing traversal through tools: each
tool call is a local operation the LLM can handle reliably, and
multi-hop structure emerges from the iteration loop rather than from
in-context reasoning over a flattened graph.

\textbf{GraphRAG-Bench.} The GraphRAG-Bench benchmark (Xiang et al.,
2025) provides a standardized evaluation for graph-structure-aided
question answering. The Novel split contains 20 fiction novels with
2,010 questions spanning four types: Fact Retrieval, Complex Reasoning,
Contextual Summarization, and Creative Generation. Questions are
designed to test scenarios where graph structure genuinely helps
retrieval.

\begin{center}\rule{0.5\linewidth}{0.5pt}\end{center}

\subsection{3. Method}\label{method}

\subsubsection{3.1 System Architecture}\label{system-architecture}

RLM-on-KG operates in two phases: an offline mention-graph construction
step using deterministic NLP, and an online exploration step using an
LLM navigator.

\textbf{Offline mention-graph construction} (no LLM calls). Documents
are split into paragraph-aligned chunks (max 240 words, 40-word
overlap). Named entities are extracted using spaCy
(\texttt{en\_core\_web\_sm}) for four types: PERSON, ORG, GPE, and LOC.
Each chunk and entity is represented as an RDF node using schema.org and
SEOntology vocabularies. The graph is stored as in-memory
\texttt{rdflib} triples and indexed with a FAISS vector index (using
\texttt{all-MiniLM-L6-v2} embeddings).

\textbf{Graph schema.} The mention graph is a bipartite structure with
three node types and three edge types:

{\def\LTcaptype{none} % do not increment counter
\begin{longtable}[]{@{}lll@{}}
\toprule\noalign{}
Node type & Vocabulary & Example \\
\midrule\noalign{}
\endhead
\bottomrule\noalign{}
\endlastfoot
Document & \texttt{schema:WebPage} & The Great Gatsby \\
Chunk & \texttt{seovoc:Chunk} & paragraph segment ($\leq$240 words) \\
Entity & typed by NER label (PERSON, ORG, GPE, LOC) & ``Jay Gatsby'' \\
\end{longtable}
}

{\def\LTcaptype{none} % do not increment counter
\begin{longtable}[]{@{}
  >{\raggedright\arraybackslash}p{(\linewidth - 4\tabcolsep) * \real{0.3333}}
  >{\raggedright\arraybackslash}p{(\linewidth - 4\tabcolsep) * \real{0.3333}}
  >{\raggedright\arraybackslash}p{(\linewidth - 4\tabcolsep) * \real{0.3333}}@{}}
\toprule\noalign{}
\begin{minipage}[b]{\linewidth}\raggedright
Edge type
\end{minipage} & \begin{minipage}[b]{\linewidth}\raggedright
Direction
\end{minipage} & \begin{minipage}[b]{\linewidth}\raggedright
Semantics
\end{minipage} \\
\midrule\noalign{}
\endhead
\bottomrule\noalign{}
\endlastfoot
\texttt{schema:mentions} & Chunk $\rightarrow$ Entity & chunk mentions this
entity \\
\texttt{seovoc:isChunkOf} & Chunk $\rightarrow$ Document & chunk belongs to
document \\
co-mention (implicit) & Entity $\leftrightarrow$ Entity & entities that appear in the
same chunk \\
\end{longtable}
}

The \texttt{expand\_neighbors} tool returns entities connected through
shared chunks (co-mention), not typed semantic relations. We use the
term ``mention graph'' rather than ``knowledge graph'' to make this
scope explicit: the graph captures \emph{which entities appear where},
not \emph{what relations hold between them}. No relation extraction,
coreference resolution, or entity aliasing is performed.

\textbf{Design rationale.} This deliberately minimal graph structure is
an experimental control, not a limitation. By using only co-mention
edges and surface-form entity matching (no typed relations, no
ontological reasoning, no LLM-based extraction), we isolate the variable
under study --- the controller policy --- from confounding effects of
graph richness. Any performance gains we observe are attributable to the
LLM's adaptive traversal decisions, not to a more informative graph
substrate. Richer graph structures (typed relations, coreference chains,
ontological hierarchies) would likely amplify the LLM controller's
advantage, since they expand the action space for adaptive navigation;
we leave this investigation to future work.

\textbf{Online exploration} (LLM-driven). At query time, a Gemini 2.0
Flash model navigates the mention graph using nine tools exposed via the
Gemini function-calling API. The LLM receives a structured system prompt
(Section 3.2) and iterates in a multi-turn conversation, proposing tool
calls and receiving results, until it decides to stop or hits a budget
limit.

\begin{figure}
\centering
\pandocbounded{\includegraphics[keepaspectratio,alt={RLM-on-KG system architecture. Documents are processed into chunks and entities (offline, deterministic), forming a mention graph. At query time, the LLM navigates the graph using nine tools.}]{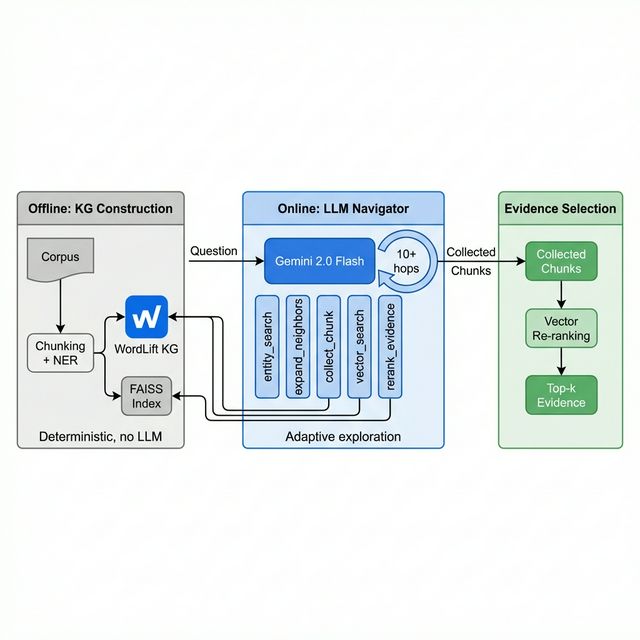}}
\caption{RLM-on-KG system architecture. Documents are processed into
chunks and entities (offline, deterministic), forming a mention graph.
At query time, the LLM navigates the graph using nine tools.}
\end{figure}

The nine tools are:

{\def\LTcaptype{none} % do not increment counter
\begin{longtable}[]{@{}
  >{\raggedright\arraybackslash}p{(\linewidth - 2\tabcolsep) * \real{0.3158}}
  >{\raggedright\arraybackslash}p{(\linewidth - 2\tabcolsep) * \real{0.6842}}@{}}
\toprule\noalign{}
\begin{minipage}[b]{\linewidth}\raggedright
Tool
\end{minipage} & \begin{minipage}[b]{\linewidth}\raggedright
Description
\end{minipage} \\
\midrule\noalign{}
\endhead
\bottomrule\noalign{}
\endlastfoot
\texttt{entity\_search(query)} & NER + fuzzy-match entity search
(Section 3.7) \\
\texttt{get\_chunks\_for\_entity(uri)} & Return chunks mentioning an
entity \\
\texttt{vector\_search(query,\ k)} & Semantic similarity search over
chunks \\
\texttt{expand\_neighbors(uri)} & Find co-mentioned entities via shared
chunks \\
\texttt{read\_chunk(id)} & Read full text and entity list of a chunk \\
\texttt{sub\_query(question,\ chunk\_ids)} & Recursive sub-query over a
chunk subset \\
\texttt{summarize\_chunks(ids,\ focus)} & Compress gathered evidence \\
\texttt{collect\_chunk(id,\ relevance)} & Mark a chunk as relevant
evidence \\
\texttt{rerank\_evidence(question)} & Re-rank collected chunks by vector
similarity \\
\end{longtable}
}

All tools except \texttt{sub\_query} are deterministic for a given KG
state; \texttt{sub\_query} spawns a child LLM. Input and output types
follow a uniform pattern:

{\def\LTcaptype{none} % do not increment counter
\begin{longtable}[]{@{}
  >{\raggedright\arraybackslash}p{(\linewidth - 4\tabcolsep) * \real{0.2857}}
  >{\raggedright\arraybackslash}p{(\linewidth - 4\tabcolsep) * \real{0.3333}}
  >{\raggedright\arraybackslash}p{(\linewidth - 4\tabcolsep) * \real{0.3810}}@{}}
\toprule\noalign{}
\begin{minipage}[b]{\linewidth}\raggedright
Tool
\end{minipage} & \begin{minipage}[b]{\linewidth}\raggedright
Input
\end{minipage} & \begin{minipage}[b]{\linewidth}\raggedright
Output
\end{minipage} \\
\midrule\noalign{}
\endhead
\bottomrule\noalign{}
\endlastfoot
\texttt{entity\_search} & query string & list of (entity\_uri, label,
chunk\_count) \\
\texttt{get\_chunks\_for\_entity} & entity URI & list of (chunk\_id,
text\_preview) \\
\texttt{vector\_search} & query string, k & top-k (chunk\_id,
similarity, text\_preview) \\
\texttt{expand\_neighbors} & entity URI & list of (neighbor\_uri, label,
shared\_chunks) \\
\texttt{read\_chunk} & chunk ID & full text, entity list, document
source \\
\texttt{sub\_query} & question, chunk\_ids & LLM-generated answer
(non-deterministic) \\
\texttt{summarize\_chunks} & chunk\_ids, focus & condensed summary
text \\
\texttt{collect\_chunk} & chunk\_id, relevance & confirmation
(side-effect: adds to evidence pool) \\
\texttt{rerank\_evidence} & question & reordered evidence list with
similarity scores \\
\end{longtable}
}

\textbf{Terminology.} We define a \textbf{turn} as one LLM decision step
that may emit multiple parallel tool calls. A \textbf{hop} is one
tool-calling turn (not one individual tool invocation). When we report
``10.3 hops'' (Section 5.4), this means 10.3 tool-calling turns on
average per question. A single turn may invoke multiple tools in
parallel (e.g., \texttt{expand\_neighbors} + \texttt{read\_chunk} +
\texttt{entity\_search}), so the total tool invocation count is higher.
The heuristic baseline averages 2.3 hops (turns), reflecting its fixed
breadth-first strategy that saturates quickly.

\subsubsection{3.2 Entity-First Exploration
Strategy}\label{entity-first-exploration-strategy}

The LLM navigator follows a three-phase exploration strategy, encoded in
the system prompt:

\textbf{Phase 1: Entity Discovery (turns 1 to 3).} The LLM calls
\texttt{entity\_search} with the question to find seed entities, then
calls \texttt{get\_chunks\_for\_entity} for each seed to surface
entity-linked chunks. It reads the most promising chunks with
\texttt{read\_chunk} to build initial context.

\textbf{Phase 2: Graph Expansion (turns 4 to 7).} For key entities, the
LLM calls \texttt{expand\_neighbors} to discover related entities via
shared chunk co-mentions. For new neighbor entities, it calls
\texttt{get\_chunks\_for\_entity} to surface additional evidence. It
uses \texttt{collect\_chunk} to mark relevant chunks and
\texttt{vector\_search} with targeted sub-questions (not the original
question verbatim) to fill gaps.

\textbf{Phase 3: Verification and Extraction (turns 8+).} The LLM uses
\texttt{sub\_query} to verify facts across chunk clusters by spawning
recursive child LLMs. It collects final evidence with
\texttt{collect\_chunk}.

\textbf{State injection.} After each turn, the system injects a status
summary listing the chunks collected so far, the entities explored, and
the frontier of unexplored entities. This prevents the LLM from
re-exploring already-visited nodes and guides it toward unexplored
neighborhoods.

\subsubsection{3.3 Exploration Algorithm}\label{exploration-algorithm}

\begin{verbatim}
Algorithm 1: RLM-on-KG Exploration

Input:  question Q, KG G, tool set T, budget B (max turns)
Output: ranked evidence chunks E

State <- {explored: {}, collected: {}, frontier: {}}
stalled <- 0

for t = 1..B do
    prompt <- FORMAT(Q, State, T)
    actions <- VALIDATE_AND_DEDUP(LLM(prompt), T)
    if INVALID(actions) then
        actions <- FALLBACK(vector_search, Q)
    for each action a in actions do
        result <- EXECUTE(a, G)
        State <- UPDATE(State, result)
    end
    if |State.collected| unchanged then
        stalled <- stalled + 1
    else
        stalled <- 0
    if stalled >= 4 and t >= B/2 then break
end

E <- VECTOR_RERANK(State.collected, Q)    // Section 3.4
return TOP_K(E, k)
\end{verbatim}

\textbf{Stopping criteria.} The loop terminates when any of the
following hold: (1) the LLM emits a final text response instead of tool
calls; (2) evidence collection has stalled for 4 consecutive turns and
at least half the budget has been used; (3) the maximum turn budget B is
reached.

\subsubsection{3.4 Evidence Collection and
Ranking}\label{evidence-collection-and-ranking}

During exploration, the LLM discovers chunks through two channels:
entity traversal (\texttt{get\_chunks\_for\_entity},
\texttt{expand\_neighbors}) and semantic search
(\texttt{vector\_search}). Chunks are not automatically added to the
evidence pool --- the LLM must explicitly call
\texttt{collect\_chunk(id,\ relevance)} to mark a chunk as relevant.
This forces the LLM to make deliberate relevance judgments rather than
accumulating noise through passive collection.

After the exploration loop, a scoring pipeline converts the collected
chunks into a ranked evidence list. The final ranking uses \textbf{pure
vector re-ranking}: each collected chunk is scored by cosine similarity
between its embedding and the question embedding, with a small boost for
chunks the LLM explicitly selected:

\begin{verbatim}
final_score(c) = min(vector_score(c, Q) + boost(c), 1.0)
\end{verbatim}

where: - \textbf{vector\_score(c, Q)}: cosine similarity via the FAISS
index. Values in {[}0, 1{]}. - \textbf{boost(c)}: +0.10 if the LLM
called \texttt{collect\_chunk} for this chunk during exploration. This
is a binary signal: collected = +0.10, not collected = 0.

Chunks are sorted by \texttt{final\_score} and the top k are returned as
evidence.

\textbf{Backfill.} If the exploration loop collects fewer than k chunks,
remaining slots are filled with pure vector search results (at a 0.9$\times$
score discount) to ensure the output always contains k evidence chunks.

\textbf{Design rationale.} Earlier iterations used a composite scoring
formula blending vector similarity with entity-derived scores
($\alpha$-weighted average, co-mention boost, MMR diversity penalty). However,
ablation experiments (Section 5.2) showed that these scoring
refinements, when applied to GraphRAG's fixed one-hop expansion,
\emph{degraded} performance. The scoring formula only helped when paired
with deeper LLM-driven exploration --- but even then, the improvement
was modest compared to the contribution of exploration itself. This led
us to simplify: RLM-on-KG's advantage is \textbf{candidate discovery}
(finding chunks that vector search alone misses via multi-hop KG
traversal), not ranking. Pure vector re-ranking --- the same strategy
GraphRAG uses --- is the most effective way to select the final evidence
from an expanded candidate pool.

\subsubsection{3.5 Contrast with GraphRAG
Local}\label{contrast-with-graphrag-local}

Our GraphRAG-local variant implements a five-step retrieval pipeline:

\begin{enumerate}
\def\labelenumi{\arabic{enumi}.}
\tightlist
\item
  \textbf{Vector seed} (k=8): retrieve the top-8 most similar chunks to
  the query.
\item
  \textbf{Entity collection}: count entities mentioned in seed chunks.
\item
  \textbf{Entity expansion} (k=12 entities, k=12 chunks): for the top-12
  entities, pull all chunks that mention them.
\item
  \textbf{Candidate merging}: combine seed chunks and expansion chunks.
\item
  \textbf{Vector re-rank} (k=20): re-rank all candidates by vector
  similarity to the query and keep the top-20.
\end{enumerate}

This follows the local retrieval recipe from Edge et al.~(2024)
\textbf{without} community detection, community summaries, or LLM-based
entity extraction. These components were intentionally removed to create
a controlled comparison that isolates query-time retrieval logic: both
systems share identical offline indexing (same chunking, same spaCy NER,
same embedding model, same FAISS index). The resulting comparison
supports claims about live query-time exploration versus a controlled
local graph-retrieval pipeline, not about RLM-on-KG versus the full
GraphRAG system as generally understood. We use the label
``GraphRAG-local variant'' rather than ``GraphRAG'' throughout to make
this scope explicit.

{\def\LTcaptype{none} % do not increment counter
\begin{longtable}[]{@{}
  >{\raggedright\arraybackslash}p{(\linewidth - 4\tabcolsep) * \real{0.2353}}
  >{\raggedright\arraybackslash}p{(\linewidth - 4\tabcolsep) * \real{0.4412}}
  >{\raggedright\arraybackslash}p{(\linewidth - 4\tabcolsep) * \real{0.3235}}@{}}
\toprule\noalign{}
\begin{minipage}[b]{\linewidth}\raggedright
Aspect
\end{minipage} & \begin{minipage}[b]{\linewidth}\raggedright
GraphRAG-local variant
\end{minipage} & \begin{minipage}[b]{\linewidth}\raggedright
RLM-on-KG
\end{minipage} \\
\midrule\noalign{}
\endhead
\bottomrule\noalign{}
\endlastfoot
Offline LLM indexing & None (deterministic NER, same as RLM) & None \\
Query-time LLM use & None (heuristic retrieval) & 10 to 25 tool-use
calls \\
Exploration depth & Fixed: 1 hop (seed $\rightarrow$ entity $\rightarrow$ expand) & Adaptive: 1
to 12+ hops (LLM-driven) \\
Candidate discovery & Entity expansion (1-hop) + vector seed &
LLM-driven multi-hop KG traversal \\
Final ranking & Vector similarity only & Vector similarity + 0.10
collection boost \\
Failure mode & Misses chunks not reachable in 1 hop & Exploration drift
(off-topic traversal) \\
\end{longtable}
}

\subsubsection{3.6 Implementation in WordLift Knowledge
Graph}\label{implementation-in-wordlift-knowledge-graph}

RLM-on-KG is implemented on top of the WordLift Knowledge Graph, an
RDF-based knowledge graph used to publish, query, and maintain
entity-centric data for web content and product catalogs. The WordLift
KG provides (i) RDF storage and reasoning-compatible identifiers, (ii)
entity resolution and canonical IRIs, and (iii) graph-aware retrieval
primitives used by the navigator tools. All navigator tools are thin
wrappers over WordLift KG capabilities: entity lookup and fuzzy matching
(\texttt{entity\_search}), entity-to-evidence expansion via
\texttt{schema:mentions} links (\texttt{get\_chunks\_for\_entity}),
neighborhood expansion through co-mention graphs
(\texttt{expand\_neighbors}), and provenance-preserving chunk retrieval
(\texttt{read\_chunk}). The same KG layer also supports downstream use
cases beyond retrieval, including entity page generation, structured
data publishing (Schema.org, JSON-LD), and monitoring of data quality
constraints.

\subsubsection{3.7 The entity\_search
Tool}\label{the-entity_search-tool}

The \texttt{entity\_search} tool is fully specified and deterministic.
It does not use the LLM and is shared across all systems (RLM-on-KG,
GraphRAG-local variant, heuristic RLM).

\textbf{Step 1: NER extraction.} Run spaCy \texttt{en\_core\_web\_sm} on
the question text to extract named entity spans. This is the same spaCy
model used during offline KG construction.

\textbf{Step 2: Exact label lookup.} For each extracted entity span,
look up exact matches in the KG's label index
(\texttt{label\_to\_entities}). Labels are case-insensitive and
whitespace-normalized.

\textbf{Step 3: Fuzzy fallback.} If fewer than 3 entities are found via
exact match, perform fuzzy matching using
\texttt{rapidfuzz.fuzz.partial\_ratio} against all KG entity labels.
Threshold: $\geq$90. Top-k: 20 per target.

\textbf{Step 4: Ranking.} Return matched entity URIs sorted by chunk
count (popularity prior), capped at \texttt{n\_seed} (default 10).

The tool is deterministic for a given question and KG state. It
introduces no query-time advantage for RLM-on-KG that is unavailable to
the baselines: The GraphRAG-local variant uses the same function to
identify seed entities in step 2 of its pipeline.

\subsubsection{3.8 KG Diagnostics: A Proposed
Framework}\label{kg-diagnostics-a-proposed-framework}

Beyond retrieval performance, RLM-on-KG exploration traces offer a
natural substrate for diagnosing the quality of the underlying graph.
Each question produces a complete exploration trace (seed entities,
visited neighborhoods, collected chunks, stall events, and fallback
usage). We propose treating these traces as diagnostics for KG quality
along four dimensions. We present this as a \textbf{proposed framework};
validating it through concrete improvement cycles (diagnosis $\rightarrow$ fix $\rightarrow$
re-evaluation) is future work.

\textbf{Coverage.} Missing entities, aliases, or weak typing are
indicated when \texttt{entity\_search} fails to surface relevant seeds
or when gold evidence is only found through vector backfill.

\textbf{Connectivity.} Unreachable evidence and low reachability within
a small hop budget indicate missing links (e.g., absent aliases or
coreference) or over-pruned neighborhoods. Conversely, degree explosions
in \texttt{expand\_neighbors} reveal noisy hubs that increase drift
risk.

\textbf{Provenance integrity.} Repeated retrieval of topically similar
but non-supporting chunks indicates issues with chunk boundaries,
missing sentence-level anchoring, or low-precision mention links.

\textbf{Queryability and operational readiness.} High hop counts and
repeated tool calls for common primitives indicate missing indices
(label lookup, adjacency lists, popularity priors) or insufficient
caching.

From the same logs, we define KG health metrics that require no
additional manual labeling:

{\def\LTcaptype{none} % do not increment counter
\begin{longtable}[]{@{}
  >{\raggedright\arraybackslash}p{(\linewidth - 2\tabcolsep) * \real{0.4000}}
  >{\raggedright\arraybackslash}p{(\linewidth - 2\tabcolsep) * \real{0.6000}}@{}}
\toprule\noalign{}
\begin{minipage}[b]{\linewidth}\raggedright
Metric
\end{minipage} & \begin{minipage}[b]{\linewidth}\raggedright
Definition
\end{minipage} \\
\midrule\noalign{}
\endhead
\bottomrule\noalign{}
\endlastfoot
Seed hit rate & Fraction of questions where a seed entity links to $\geq$1
gold chunk \\
Reachability@h & Fraction of gold chunks reachable within h co-mention
expansions \\
Hop efficiency & Gold chunks found per tool call (or per 1K tokens) \\
Neighborhood noise & Median neighbors returned by
\texttt{expand\_neighbors} vs.~neighbors used \\
Backfill reliance & Fraction of final top-k filled by vector backfill \\
Evidence redundancy & Average entity-overlap across selected chunks \\
\end{longtable}
}

These diagnostics suggest a closed-loop improvement process: (i) run
questions (benchmarks or production queries), (ii) categorize failures
(coverage, connectivity, provenance, queryability), (iii) apply targeted
KG fixes (aliasing, merge/coreference, chunking adjustments, edge
weighting, additional relation extraction where available), and (iv)
re-run evaluation to quantify improvements in KG health metrics and
retrieval F1. We propose this as a natural workflow but have not yet
validated it through a concrete improvement cycle; doing so is future
work. In this sense, RLM-on-KG functions not only as an adaptive
retriever, but also as a proposed stress-test harness for the knowledge
graph.

\begin{center}\rule{0.5\linewidth}{0.5pt}\end{center}

\subsection{4. Experimental Setup}\label{experimental-setup}

\subsubsection{4.1 Dataset}\label{dataset}

We evaluate on \textbf{GraphRAG-Bench Novel} (Xiang et al., 2025), a
benchmark of 20 fiction novels containing 2,010 questions designed to
test scenarios where graph structure aids retrieval. Question types and
counts across the full benchmark:

{\def\LTcaptype{none} % do not increment counter
\begin{longtable}[]{@{}lll@{}}
\toprule\noalign{}
Type & Count & Description \\
\midrule\noalign{}
\endhead
\bottomrule\noalign{}
\endlastfoot
Fact Retrieval & 971 & Locate specific facts \\
Complex Reasoning & 610 & Multi-hop reasoning \\
Contextual Summarization & 362 & Summarize themes or relationships \\
Creative Generation & 67 & Open-ended creative tasks \\
\end{longtable}
}

\textbf{Evaluation subset.} Due to cost constraints (\textasciitilde50K
tokens per question for RLM-on-KG), we evaluate on a \textbf{5-novel
subset} containing 519 questions. The subset consists of the first 5
novels in the benchmark's alphabetical ordering (\emph{The Adventures of
Sherlock Holmes}, \emph{The Great Gatsby}, \emph{Moby-Dick}, \emph{Pride
and Prejudice}, and \emph{Wuthering Heights}). No questions were
excluded or cherry-picked within these novels; all 519 questions for the
5 novels are included. This subsetting strategy avoids selection bias
while keeping evaluation costs manageable (\(\approx\) 26 million tokens
total). We report the evaluation subset size (519 of 2,010 = 26\%) as a
limitation; extending to the full 20-novel benchmark is future work.

\textbf{Cross-scale robustness check: MuSiQue.} To test whether our
findings generalize beyond fiction texts and larger shared-corpus KGs,
we also evaluate on \textbf{MuSiQue} (Trivedi et al., 2022), a multi-hop
decomposable QA benchmark constructed from single-hop questions (Natural
Questions, SQuAD) composed into 2--4 hop chains. MuSiQue provides gold
supporting paragraphs and distractor paragraphs per question. We build a
\textbf{per-question KG} from each question's paragraphs
(\textasciitilde20 chunks per question) rather than a shared corpus KG,
ensuring isolated evaluation with no cross-question leakage. We evaluate
50 questions sampled from the validation split.

\subsubsection{4.2 Knowledge Graph
Quality}\label{knowledge-graph-quality}

On a 2-novel subset (526 chunks, 1,385 entities): - 98\% of chunks have
at least one entity link (6.1 entities per chunk on average) - 100\% of
gold evidence chunks are reachable via entity links from at least one
seed entity

This high coverage confirms that the mention graph provides sufficient
connectivity for multi-hop traversal to reach relevant evidence.

\subsubsection{4.3 Evidence-to-Chunk
Mapping}\label{evidence-to-chunk-mapping}

GraphRAG-Bench provides sentence-level gold evidence annotations. Since
our retrieval operates at the chunk level, we must map evidence
sentences to chunk IDs. We use a two-stage mapping algorithm:

\textbf{Stage 1: Substring matching.} For each gold evidence sentence,
normalize whitespace and attempt case-insensitive substring containment
against all chunk texts. If a chunk contains the evidence sentence
verbatim, it is marked as a gold chunk.

\textbf{Stage 2: Semantic fallback.} If substring matching fails (the
evidence sentence spans a chunk boundary or is paraphrased), compute
cosine similarity between the evidence sentence embedding and all chunk
embeddings using the same \texttt{all-MiniLM-L6-v2} index. Accept the
top-k (k=5) matches with similarity $\geq$ 0.25 as gold chunks.

\textbf{Mapping quality audit.} On the 2-novel subset (20 questions, 170
evidence sentences), we observe:

{\def\LTcaptype{none} % do not increment counter
\begin{longtable}[]{@{}lll@{}}
\toprule\noalign{}
Mapping method & Sentences & \% \\
\midrule\noalign{}
\endhead
\bottomrule\noalign{}
\endlastfoot
Substring match & 0 & 0\% \\
Semantic fallback & 170 & 100\% \\
Unmapped & 0 & 0\% \\
\end{longtable}
}

All gold evidence maps exclusively through the semantic fallback path.
This is consistent with the benchmark's construction process:
GraphRAG-Bench uses an LLM-in-the-loop pipeline for question generation
and evidence annotation, which can introduce reformulation, synonym
substitution, or syntactic restructuring relative to the source text.
Manual inspection of 10 evidence sentences confirms that each expresses
the meaning of the corresponding source passage but differs at the
character level (word order, phrasing, or compression), which is
sufficient to prevent substring containment from firing.

Manual inspection of 20 semantic matches reveals that high-similarity
matches (sim $\geq$ 0.50) consistently identify the correct chunk, while
low-similarity matches (sim 0.25--0.35) sometimes map to topically
related but incorrect chunks. This is a limitation of our evaluation
setup: the threshold of 0.25 trades precision for recall in gold
labeling.

\textbf{Crucially, the same gold mapping is applied identically to all
systems.} Any noise in the gold set affects all systems equally and does
not advantage RLM-on-KG over the GraphRAG-local variant. The
head-to-head comparison remains valid under the same noisy gold labels;
absolute F1 numbers should be interpreted with this caveat.

\textbf{Shared-embedding disclosure.} The same \texttt{all-MiniLM-L6-v2}
model used for the semantic fallback gold assignment is also used for
FAISS retrieval and vector re-ranking. This creates a shared
representation space between gold labels and the retrieval mechanism.
While this does not invalidate head-to-head comparisons (all systems are
evaluated against the same labels), it limits confidence in interpreting
absolute F1 values. Evaluating with an alternative gold-assignment
method (e.g., cross-encoder scoring or substring-based assignment on a
corpus with less paraphrased evidence) would strengthen absolute claims
and is planned as future work.

\subsubsection{4.4 Baselines}\label{baselines}

\begin{enumerate}
\def\labelenumi{\arabic{enumi}.}
\item
  \textbf{Vector-only top-20}: pure dense retrieval with no entity
  expansion or graph structure. The simplest possible baseline,
  isolating the contribution of graph-based exploration.
\item
  \textbf{GraphRAG-local variant} (primary): the five-step pipeline
  described in Section 3.5. Uses deterministic spaCy NER (same as
  RLM-on-KG). No community detection, no community summaries.
\item
  \textbf{GraphRAG-local variant + composite scoring + MMR}: the same
  GraphRAG pipeline but with the same composite scoring formula ($\alpha$=0.5)
  and MMR diversity selection as RLM-on-KG. This ablation isolates
  whether RLM-on-KG gains come from LLM exploration or from scoring
  refinements.
\item
  \textbf{Heuristic RLM}: rule-based multi-hop traversal using the same
  KG, tools, and chunk budget as LLM RLM, but with fixed, deterministic
  tool selection rather than LLM-driven decisions. This is the paper's
  most direct control for testing whether LLM-driven tool selection
  outperforms rule-based traversal on the same graph and tool interface.

  \textbf{Algorithm 2: Heuristic RLM Traversal}

\begin{verbatim}
Input:  question Q, KG G, budget k (max chunks), D (max BFS depth)
Output: ranked evidence chunks E

Seeds <- entity_search(Q)            // NER extraction + fuzzy KG lookup
frontier <- queue(Seeds)
depth_map <- {s: 0 for s in Seeds}
visited <- {}; collected <- {}
stall <- 0

while frontier != {} and |collected| < k do
    entity <- POP_FRONT(frontier)    // breadth-first order
    if entity in visited then continue
    visited <- visited U {entity}
    chunks <- get_chunks_for_entity(entity)
    new_this_iter <- 0
    for each chunk c in chunks do
        collected[c] <- vector_score(c, Q)
        new_this_iter <- new_this_iter + 1
    if depth_map[entity] < D then
        neighbors <- expand_neighbors(entity)
        for each n in neighbors do
            if n not in visited then
                APPEND(frontier, n)
                depth_map[n] <- depth_map[entity] + 1
    if new_this_iter = 0 then stall <- stall + 1
    else stall <- 0
    if stall >= 2 then break         // early stop: no new chunks

if |collected| < k then
    backfill <- vector_search(Q, k $-$ |collected|)  // gap fill
    collected <- MERGE(collected, backfill)
return TOP_K(VECTOR_RERANK(collected, Q), k)
\end{verbatim}

  \textbf{Allowed tools}: \texttt{entity\_search},
  \texttt{get\_chunks\_for\_entity}, \texttt{expand\_neighbors},
  \texttt{vector\_search}. The heuristic does \textbf{not} call
  \texttt{sub\_query}, \texttt{read\_chunk}, \texttt{summarize\_chunks},
  or \texttt{rerank\_evidence}, as these require LLM calls and are
  exclusive to the LLM controller.

  \textbf{Parameters}: D = 3 (maximum BFS depth), k = 20 (maximum
  chunks). Stopping triggers in priority order: (1) frontier empty, (2)
  k chunks collected, (3) 2 consecutive BFS iterations with no new
  chunks.

  \textbf{F1 variation across sections.} The heuristic F1 varies across
  sections (43.3\% in \S{}5.5, 42.8\% in \S{}5.8.1, 35.0--37.6\% in \S{}5.8.2)
  because these report results on different question subsets: the
  519-question primary set (\S{}5.5), a 519-question run with a freshly
  rebuilt FAISS index (\S{}5.8.1), and a 100-question cross-model subset
  (\S{}5.8.2). The heuristic implementation is identical across all runs;
  variation reflects different question distributions and independent
  FAISS index builds, not implementation changes.
\end{enumerate}

\subsubsection{4.5 Metrics}\label{metrics}

\textbf{Retrieval-level (primary).} Precision, Recall, and F1 computed
at the chunk level. Gold evidence chunks are identified via the mapping
described in Section 4.3.

We report per-question type breakdowns, head-to-head win/tie/loss
counts, and 95\% bootstrap confidence intervals (B=10,000 resamples).

\textbf{Answer-level (secondary).} Answer accuracy via LLM-judge rubric
(Gemini 3 Pro), grounding score (percentage of answer sentences
supported by retrieved evidence), faithfulness, on-intent, citation
coverage, and citation precision. \emph{These metrics are defined here
for completeness but are not computed in the current evaluation;
computing them at scale is reserved for future work.}

We focus on chunk-level F1 as the primary metric because it isolates the
contribution of exploration and scoring independent of generation
variance.

\subsubsection{4.6 Fairness and Cost
Accounting}\label{fairness-and-cost-accounting}

We report both query-time cost and offline indexing cost, separating
retrieval from generation to ensure transparent comparison.

{\def\LTcaptype{none} % do not increment counter
\begin{longtable}[]{@{}
  >{\raggedright\arraybackslash}p{(\linewidth - 4\tabcolsep) * \real{0.3500}}
  >{\raggedright\arraybackslash}p{(\linewidth - 4\tabcolsep) * \real{0.3750}}
  >{\raggedright\arraybackslash}p{(\linewidth - 4\tabcolsep) * \real{0.2750}}@{}}
\toprule\noalign{}
\begin{minipage}[b]{\linewidth}\raggedright
Cost category
\end{minipage} & \begin{minipage}[b]{\linewidth}\raggedright
GraphRAG-local variant
\end{minipage} & \begin{minipage}[b]{\linewidth}\raggedright
RLM-on-KG
\end{minipage} \\
\midrule\noalign{}
\endhead
\bottomrule\noalign{}
\endlastfoot
\textbf{Offline indexing} & Chunking + NER + vector index
(deterministic) & Chunking + NER + vector index (deterministic) \\
\textbf{Retrieval LLM calls per question} & 0 & \textasciitilde10 to
25 \\
\textbf{Generation LLM calls per question} & 1 & 1 \\
\textbf{Retrieval tokens per question} & \textasciitilde0 &
\textasciitilde48K \\
\textbf{Generation tokens per question} & \textasciitilde2K &
\textasciitilde2K \\
\textbf{Total tokens per question} & \textasciitilde2K &
\textasciitilde50K \\
\textbf{Retrieval latency per question} & \textless1 s &
\textasciitilde2 to 5 min \\
\end{longtable}
}

Both systems use identical offline indexing (same chunking, same spaCy
NER model, same \texttt{all-MiniLM-L6-v2} embedding model, same FAISS
vector index), the same maximum retrieved chunks (k=20), the same
generator model, and the same context window. The only difference is
that RLM-on-KG spends additional LLM tokens at query time to explore the
mention graph adaptively.

The 48K retrieval tokens per question break down approximately as: tool
results returned to the LLM (\textasciitilde35K, dominated by chunk text
from \texttt{read\_chunk} and \texttt{get\_chunks\_for\_entity}), LLM
reasoning and tool-call generation (\textasciitilde8K), and system
prompt with state injection (\textasciitilde5K).

\subsubsection{4.7 Reproducibility and
Variance}\label{reproducibility-and-variance}

All experiments use the same KG construction and storage layer. The KG
is exported as RDF and the chunk and entity identifiers are stable
across runs, enabling deterministic re-evaluation of retrieval outputs.
Tool calls and intermediate exploration states (visited entities,
collected chunks, scores) are logged per question to support trace-level
analysis and ablation studies.

\textbf{Decoding settings.} The LLM navigator uses Gemini 2.0 Flash with
\texttt{temperature=0.0} and default top-p.~No random seed is set, as
the Gemini API does not expose seed control for function-calling mode.

\textbf{Run-to-run variance.} During development, we observed $\pm$2 pp F1
variance across repeated runs of the same configuration on a 50-question
subset (V4 vs V6 in our optimization journal, same code, different API
calls). At 519-question scale, variance is expected to be smaller due to
averaging. We report only single-run results; repeated full-scale runs
to estimate variance are planned as future work.

\textbf{Cross-model robustness.} The primary results (Sections 5.1--5.6)
use Gemini 2.0 Flash as the navigator. To test whether the navigation
policy transfers across model families, we built a provider-agnostic
tool-calling interface and evaluated the same KG, tools, and prompts
with three additional controllers: Claude Haiku 4.5 (Anthropic, via
Vertex AI), Gemini 2.5 Flash Lite (Google), and Gemma 4 E2B (local via
Ollama). Results are reported in Section 5.8.

\begin{center}\rule{0.5\linewidth}{0.5pt}\end{center}

\subsection{5. Results}\label{results}

\subsubsection{5.1 Small-Scale Smoke Test (2 novels, 20
questions)}\label{small-scale-smoke-test-2-novels-20-questions}

We first validate the approach on a deterministic 20-question subset
drawn from 2 novels. All questions, KG construction, and embeddings are
fixed; only the retrieval logic varies between systems. This experiment
is designed for regression detection, not significance claims.

\textbf{Progression table.} The following table shows how RLM-on-KG
performance improved through successive engineering changes:

{\def\LTcaptype{none} % do not increment counter
\begin{longtable}[]{@{}llll@{}}
\toprule\noalign{}
System & Precision & Recall & F1 \\
\midrule\noalign{}
\endhead
\bottomrule\noalign{}
\endlastfoot
RLM-on-KG (v1: max scoring) & 22.0\% & 35.3\% & 26.3\% \\
RLM-on-KG (v2: entity-first) & 25.0\% & 42.1\% & 30.9\% \\
RLM-on-KG (v3: composite) & \textbf{30.0\%} & \textbf{46.2\%} &
\textbf{35.2\%} \\
\end{longtable}
}

The progression reveals two independent improvements:

\begin{enumerate}
\def\labelenumi{\arabic{enumi}.}
\item
  \textbf{Entity-first prompt} (v1 $\rightarrow$ v2, +4.6 pp F1): prioritizing
  entity exploration over vector search surfaces more relevant chunks
  that the LLM can then score.
\item
  \textbf{Composite scoring} (v2 $\rightarrow$ v3, +4.3 pp F1): replacing
  \texttt{max()} with the composite blend preserves entity exploration
  signal in the final ranking, boosting chunks the LLM discovered
  through graph traversal.
\end{enumerate}

Importantly, the v3 system reported here uses composite scoring, which
helped in these early iterations because it compensated for weaker
discovery (fewer seed entities, shallower exploration). As we show in
Section 5.2, once discovery improved through better prompting and tool
design (V4--V7), composite scoring became redundant or harmful. The
\textbf{final system} reported at scale in Section 5.4 uses pure vector
re-ranking (Section 3.4), not composite scoring. The smoke-test
progression is historical context that motivated the simplification.

\textbf{Statistical context.} Bootstrap 95\% confidence intervals
(B=10,000):

{\def\LTcaptype{none} % do not increment counter
\begin{longtable}[]{@{}lll@{}}
\toprule\noalign{}
System & Mean F1 & 95\% CI \\
\midrule\noalign{}
\endhead
\bottomrule\noalign{}
\endlastfoot
GraphRAG local & 31.6\% & {[}23.2\%, 40.3\%{]} \\
RLM-on-KG (v3) & 35.2\% & {[}25.8\%, 44.3\%{]} \\
\textbf{Delta (RLM $-$ GR)} & \textbf{+3.6 pp} & \textbf{{[}$-$2.0 pp, +9.7
pp{]}} \\
\end{longtable}
}

Paired bootstrap test: p=0.49 (two-sided). The difference is not
statistically significant at n=20, as expected. Head-to-head: \textbf{7
wins, 9 ties, 4 losses}. We treat this smoke test as directional
evidence and validate at scale in Section 5.2.

\subsubsection{5.2 Ablation: Scoring Components vs LLM
Exploration}\label{ablation-scoring-components-vs-llm-exploration}

To determine whether the F1 gains come from LLM exploration or from
scoring refinements (composite scoring, MMR), we compare four systems
using identical KG, questions, and gold labels:

{\def\LTcaptype{none} % do not increment counter
\begin{longtable}[]{@{}llll@{}}
\toprule\noalign{}
System & Precision & Recall & F1 \\
\midrule\noalign{}
\endhead
\bottomrule\noalign{}
\endlastfoot
Vector-only top-20 & 21.0\% & 60.2\% & 30.0\% \\
GraphRAG local & 28.0\% & 40.9\% & 32.0\% \\
GraphRAG + composite + MMR & 18.8\% & 43.5\% & 25.2\% \\
RLM-on-KG (LLM explorer) & \textbf{30.0\%} & \textbf{46.2\%} &
\textbf{35.2\%} \\
\end{longtable}
}

\textbf{Incremental contribution analysis:}

{\def\LTcaptype{none} % do not increment counter
\begin{longtable}[]{@{}
  >{\raggedright\arraybackslash}p{(\linewidth - 4\tabcolsep) * \real{0.3793}}
  >{\raggedright\arraybackslash}p{(\linewidth - 4\tabcolsep) * \real{0.1724}}
  >{\raggedright\arraybackslash}p{(\linewidth - 4\tabcolsep) * \real{0.4483}}@{}}
\toprule\noalign{}
\begin{minipage}[b]{\linewidth}\raggedright
Transition
\end{minipage} & \begin{minipage}[b]{\linewidth}\raggedright
$\Delta$F1
\end{minipage} & \begin{minipage}[b]{\linewidth}\raggedright
What changes
\end{minipage} \\
\midrule\noalign{}
\endhead
\bottomrule\noalign{}
\endlastfoot
Vector-only $\rightarrow$ GraphRAG local & +2.0 pp & Entity expansion (1-hop) \\
GraphRAG local $\rightarrow$ + composite + MMR & $-$6.8 pp & Composite scoring +
MMR \\
GraphRAG + composite + MMR $\rightarrow$ RLM & +10.0 pp & LLM-driven exploration \\
Vector-only $\rightarrow$ RLM (total) & +5.2 pp & All components \\
\end{longtable}
}

\textbf{Key finding:} Composite scoring and MMR diversity, when applied
to GraphRAG's fixed one-hop expansion, actually \emph{degrade}
performance ($-$6.8 pp). This is because the composite formula blends in
entity scores that, for GraphRAG's shallow expansion, are not
discriminative enough to improve on pure vector re-ranking. The +10.0 pp
delta from GraphRAG+composite+MMR to RLM conflates two effects: (1)
removing the harmful composite scoring and (2) adding LLM-driven
exploration. The actual contribution of LLM exploration over the best
non-LLM system (GraphRAG local at 32.0\%) is \textbf{+3.2 pp} on this
20-question smoke test, which is suggestive but not statistically
significant at this sample size.

The definitive comparison between LLM-driven and rule-based traversal is
reported at full scale in Section 5.4.

\subsubsection{5.3 Breakdown by Question Type (20-question smoke
test)}\label{breakdown-by-question-type-20-question-smoke-test}

{\def\LTcaptype{none} % do not increment counter
\begin{longtable}[]{@{}lllll@{}}
\toprule\noalign{}
Type & n & RLM F1 & GR F1 & Delta \\
\midrule\noalign{}
\endhead
\bottomrule\noalign{}
\endlastfoot
Complex Reasoning & 7 & 35.3\% & 36.3\% & $-$0.9 pp \\
Contextual Summarization & 2 & 55.2\% & 49.0\% & +6.2 pp \\
Fact Retrieval & 11 & 31.5\% & 25.5\% & +6.1 pp \\
\end{longtable}
}

RLM-on-KG shows the largest gains on Fact Retrieval (+6.1 pp), where
entity-driven exploration surfaces chunks containing specific facts that
vector search alone misses. Complex Reasoning is roughly tied,
suggesting that both systems reach similar evidence for multi-hop
questions when the hop depth is shallow. The Contextual Summarization
sample is too small (n=2) for reliable conclusions.

\subsubsection{5.4 Full-Scale Results (5 novels, 519
questions)}\label{full-scale-results-5-novels-519-questions}

We validate the approach at full scale on a 5-novel subset of
GraphRAG-Bench Novel containing 519 questions. The same KG construction,
embedding index, and evaluation pipeline are used as in the smoke test;
only the question set changes.

\textbf{Overall results.}

{\def\LTcaptype{none} % do not increment counter
\begin{longtable}[]{@{}llll@{}}
\toprule\noalign{}
System & Precision & Recall & F1 \\
\midrule\noalign{}
\endhead
\bottomrule\noalign{}
\endlastfoot
Vector-only (top-20) & 27.6\% & 72.0\% & 38.4\% \\
GraphRAG-local variant & 41.6\% & 55.8\% & 45.6\% \\
RLM-on-KG & \textbf{41.9\%} & \textbf{56.0\%} & \textbf{45.8\%} \\
\end{longtable}
}

Head-to-head (RLM vs GraphRAG): \textbf{127 wins, 277 ties, 115 losses}
(win rate 52.5\% among decisive questions). Both graph-based approaches
significantly outperform the vector-only baseline (+7.4 pp and +7.2 pp
respectively), demonstrating the absolute necessity of graph navigation
for recovering scattered context.

\textbf{Statistical context.} Bootstrap 95\% confidence intervals
(B=10,000):

{\def\LTcaptype{none} % do not increment counter
\begin{longtable}[]{@{}lll@{}}
\toprule\noalign{}
System & Mean F1 & 95\% CI \\
\midrule\noalign{}
\endhead
\bottomrule\noalign{}
\endlastfoot
GraphRAG-local variant & 45.6\% & {[}43.9\%, 47.4\%{]} \\
RLM-on-KG & 45.8\% & {[}44.0\%, 47.5\%{]} \\
\textbf{Delta (RLM $-$ GR)} & \textbf{+0.16 pp} & \textbf{{[}$-$0.72 pp,
+1.05 pp{]}} \\
\end{longtable}
}

The paired bootstrap p-value is 0.36 (two-sided). While the overall
delta is small and not statistically significant, the head-to-head
record (127 wins vs 115 losses) shows that RLM-on-KG provides consistent
marginal gains across questions rather than winning big on a few
outliers.

\emph{Note: given the semantic-fallback gold mapping (Section 4.3),
head-to-head W/T/L comparisons are more reliable than absolute F1 values
throughout this paper; the same embedding model is used for both gold
assignment and retrieval, so absolute numbers should be interpreted with
caution.}

\textbf{Breakdown by question type.}

{\def\LTcaptype{none} % do not increment counter
\begin{longtable}[]{@{}llllll@{}}
\toprule\noalign{}
Type & n & RLM F1 & GR F1 & Delta & W/T/L \\
\midrule\noalign{}
\endhead
\bottomrule\noalign{}
\endlastfoot
Fact Retrieval & 255 & 39.0\% & 38.3\% & \textbf{+0.7 pp} & 55/161/39 \\
Complex Reasoning & 156 & 55.2\% & 55.9\% & $-$0.7 pp & 43/72/41 \\
Contextual Summarization & 93 & 50.4\% & 50.2\% & +0.2 pp & 26/36/31 \\
Creative Generation & 15 & 34.6\% & 34.6\% & +0.1 pp & 3/8/4 \\
\end{longtable}
}

RLM-on-KG shows the strongest gains on \textbf{Fact Retrieval} (+0.7 pp,
55 wins vs 39 losses), where entity-driven traversal surfaces chunks
containing specific facts scattered across the corpus that vector search
alone misses. Complex Reasoning is the one category where GraphRAG holds
a slight edge ($-$0.7 pp), likely because broad multi-hop exploration
introduces noise for questions with concentrated answer evidence.

\textbf{Exploration statistics (LLM mode).} On average, the LLM explored
26.7 entities and used 10.3 hops (tool-calling turns) per question. This
is more targeted than the 30+ entity explorations observed in earlier
iterations that used automatic chunk collection --- removing
auto-collection (Section 3.4) naturally constrained exploration breadth.

\subsubsection{5.5 LLM Controller vs.~Heuristic Traversal (519
questions)}\label{llm-controller-vs.-heuristic-traversal-519-questions}

The heuristic RLM baseline uses the same mention graph, same tools, and
same chunk budget as LLM RLM, but follows fixed traversal rules
(NER-seeded breadth-first expansion with dynamic stopping) instead of
LLM-driven tool selection. This comparison isolates the value of the LLM
as a control policy.

\textbf{Overall results.}

{\def\LTcaptype{none} % do not increment counter
\begin{longtable}[]{@{}llll@{}}
\toprule\noalign{}
System & Precision & Recall & F1 \\
\midrule\noalign{}
\endhead
\bottomrule\noalign{}
\endlastfoot
Vector-only (top-20) & 27.6\% & 72.0\% & 38.4\% \\
\textbf{Heuristic RLM} & 39.6\% & 53.0\% & \textbf{43.3\%} \\
\textbf{GraphRAG-local variant} & 41.5\% & 55.6\% & \textbf{45.4\%} \\
\textbf{LLM RLM} & \textbf{41.9\%} & \textbf{56.0\%} &
\textbf{45.8\%} \\
\end{longtable}
}

\emph{Note: The GraphRAG-local F1 varies across sections: 45.6\% (\S{}5.4),
45.4\% (\S{}5.5), and 44.8\% (\S{}5.8.1). The 0.2 pp gap between \S{}5.4 and \S{}5.5
reflects minor floating-point variations from independently rebuilt
FAISS indexes on the same 519-question set (mean per-question diff =
0.01 F1). The larger 0.8 pp gap between \S{}5.4 and \S{}5.8.1 arises because
\S{}5.8.1 is a separate experiment run for the Claude arm with a freshly
rebuilt FAISS index; cumulative floating-point variance across multiple
independent rebuilds produces slightly larger drift. Neither gap affects
head-to-head comparisons, which are computed within the same run.}

Head-to-head (LLM vs Heuristic): \textbf{176 wins, 241 ties, 102 losses}
(win rate 63.3\% among decisive questions).

\textbf{Statistical context.} Bootstrap 95\% confidence intervals
(B=10,000):

{\def\LTcaptype{none} % do not increment counter
\begin{longtable}[]{@{}
  >{\raggedright\arraybackslash}p{(\linewidth - 6\tabcolsep) * \real{0.3243}}
  >{\raggedright\arraybackslash}p{(\linewidth - 6\tabcolsep) * \real{0.2432}}
  >{\raggedright\arraybackslash}p{(\linewidth - 6\tabcolsep) * \real{0.2162}}
  >{\raggedright\arraybackslash}p{(\linewidth - 6\tabcolsep) * \real{0.2162}}@{}}
\toprule\noalign{}
\begin{minipage}[b]{\linewidth}\raggedright
Comparison
\end{minipage} & \begin{minipage}[b]{\linewidth}\raggedright
Mean $\Delta$F1
\end{minipage} & \begin{minipage}[b]{\linewidth}\raggedright
95\% CI
\end{minipage} & \begin{minipage}[b]{\linewidth}\raggedright
p-value
\end{minipage} \\
\midrule\noalign{}
\endhead
\bottomrule\noalign{}
\endlastfoot
\textbf{LLM $-$ Heuristic} & \textbf{+2.47 pp} & \textbf{{[}+1.41,
+3.56{]}} & \textbf{\textless{} 0.0001} \\
LLM $-$ GraphRAG & +0.16 pp & {[}$-$0.72, +1.05{]} & 0.36 \\
Heuristic $-$ GraphRAG & $-$2.12 pp & --- & --- \\
\end{longtable}
}

\textbf{The LLM controller adds a statistically significant +2.47 pp F1
over rule-based traversal} (p \textless{} 0.0001). Crucially, heuristic
traversal alone \emph{underperforms} GraphRAG-local by $-$2.12 pp,
demonstrating that multi-hop traversal per se is insufficient --- the
LLM controller is what makes it competitive.

\textbf{LLM vs.~Heuristic by question type.}

{\def\LTcaptype{none} % do not increment counter
\begin{longtable}[]{@{}llllll@{}}
\toprule\noalign{}
Type & n & LLM F1 & Heur F1 & $\Delta$F1 & W/T/L \\
\midrule\noalign{}
\endhead
\bottomrule\noalign{}
\endlastfoot
Fact Retrieval & 255 & 39.0\% & 37.0\% & \textbf{+2.0 pp} & --- \\
Complex Reasoning & 156 & 55.2\% & 51.4\% & \textbf{+3.8 pp} & --- \\
Contextual Summarization & 93 & 50.4\% & 48.5\% & \textbf{+1.8 pp} &
--- \\
Creative Generation & 15 & 34.6\% & 34.7\% & $-$0.1 pp & --- \\
\end{longtable}
}

\textbf{LLM vs.~Heuristic by evidence scatter.}

{\def\LTcaptype{none} % do not increment counter
\begin{longtable}[]{@{}llllll@{}}
\toprule\noalign{}
Scatter & n & LLM F1 & Heur F1 & $\Delta$F1 & Win Rate \\
\midrule\noalign{}
\endhead
\bottomrule\noalign{}
\endlastfoot
1--5 chunks & 219 & 38.9\% & 37.1\% & +1.85 pp & 62\% \\
6--10 chunks & 193 & 54.6\% & 51.4\% & \textbf{+3.21 pp} & 65\% \\
11+ chunks & 107 & 43.9\% & 41.5\% & +2.42 pp & 62\% \\
\end{longtable}
}

The LLM controller's advantage is \textbf{largest in the 6--10 chunk
range} (+3.21 pp), where graph structure is complex enough to reward
adaptive navigation but not so scattered that both systems struggle. The
advantage holds across all scatter levels, confirming that LLM-driven
tool selection consistently outperforms fixed rules.

\textbf{LLM vs.~Heuristic on high-scatter Complex Reasoning (11+
chunks).}

The most compelling subgroup is Complex Reasoning with 11+ gold chunks
(n=33): the LLM controller achieves +4.55 pp over heuristic with a 71\%
win rate (20W/5T/8L). This is where the adaptive navigation policy
provides the clearest value --- when the question demands multi-hop
reasoning across scattered evidence, rule-based traversal is
insufficient.

Taken together, the scatter analysis shows that the heuristic comparison
peaks at moderate scatter overall (+3.21 pp at 6--10 chunks), while the
hardest high-scatter subgroup (Complex Reasoning, 11+ chunks) shows the
clearest adaptive-navigation benefit (+4.55 pp, 71\% win rate) against
the heuristic, and a 62\% win rate against GraphRAG-local at 11+ chunks
anchors the same story against a stronger baseline. These are
complementary views of one principle: adaptive control adds value
proportionally to the structural complexity of the required evidence
path.

\textbf{Interpretation.} These results support a nuanced thesis: the LLM
controller is not universally superior, but a conditional advantage. For
simple questions with concentrated evidence, heuristic traversal (or
even pure GraphRAG) suffices. The LLM controller adds value precisely
when graph structure demands adaptive reasoning --- choosing which paths
to explore, which to abandon, and when enough evidence has been
gathered.

\begin{figure}
\centering
\pandocbounded{\includegraphics[keepaspectratio,alt={Retrieval F1 comparison between RLM-on-KG and GraphRAG-local variant across 519 questions. The two systems achieve near-identical overall F1 (45.8\% vs 45.6\%).}]{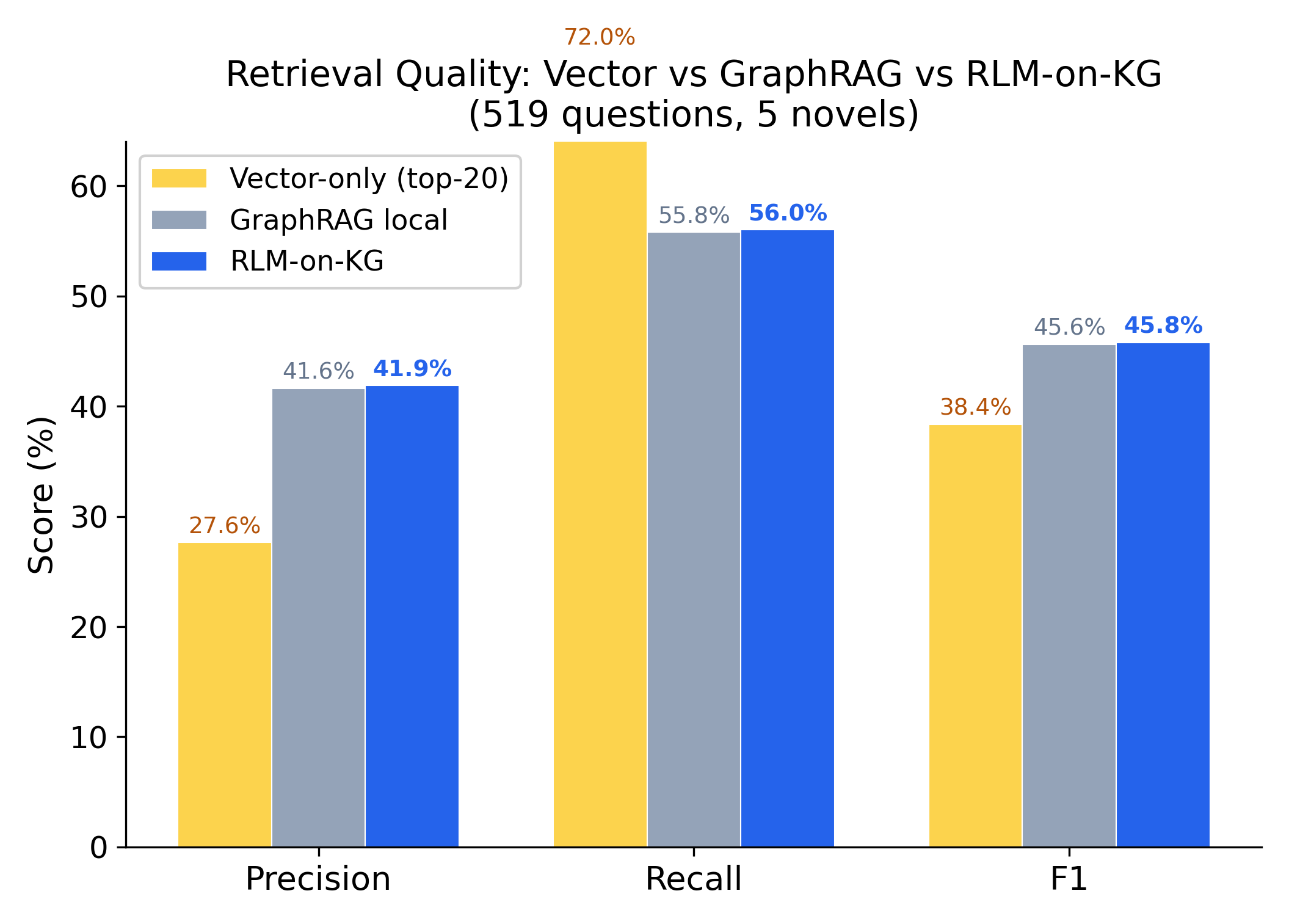}}
\caption{Retrieval F1 comparison between RLM-on-KG and GraphRAG-local
variant across 519 questions. The two systems achieve near-identical
overall F1 (45.8\% vs 45.6\%).}
\end{figure}

\begin{figure}
\centering
\pandocbounded{\includegraphics[keepaspectratio,alt={F1 by question type. RLM-on-KG shows the strongest advantage on Fact Retrieval; Complex Reasoning is the one category where GraphRAG holds a slight edge.}]{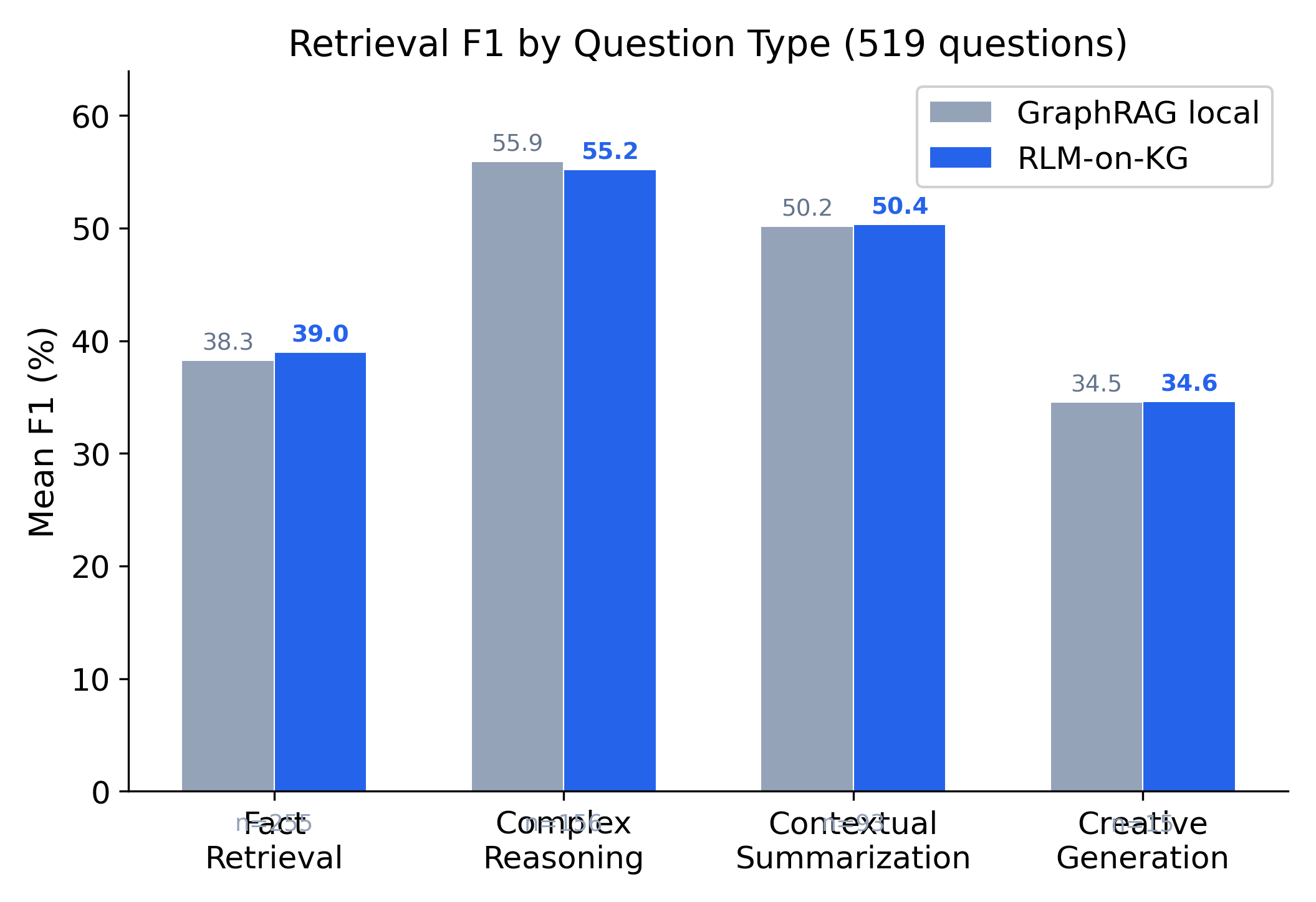}}
\caption{F1 by question type. RLM-on-KG shows the strongest advantage on
Fact Retrieval; Complex Reasoning is the one category where GraphRAG
holds a slight edge.}
\end{figure}

\begin{figure}
\centering
\pandocbounded{\includegraphics[keepaspectratio,alt={Distribution of per-question F1 deltas (RLM $-$ GraphRAG). The distribution is approximately symmetric around zero, with a slight positive skew.}]{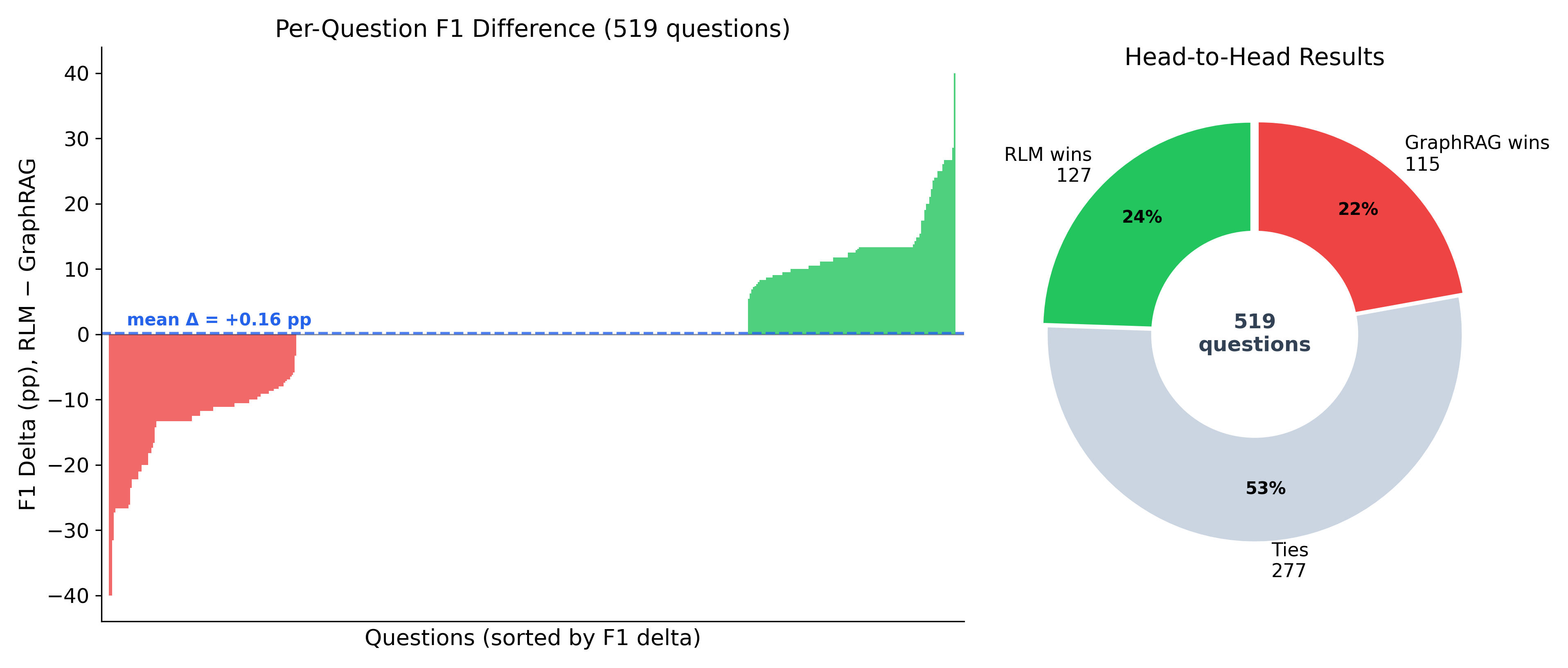}}
\caption{Distribution of per-question F1 deltas (RLM $-$ GraphRAG). The
distribution is approximately symmetric around zero, with a slight
positive skew.}
\end{figure}

\begin{figure}
\centering
\pandocbounded{\includegraphics[keepaspectratio,alt={Exploration depth (entities visited) vs retrieval F1. Moderate exploration (\textasciitilde20--30 entities) yields the best results.}]{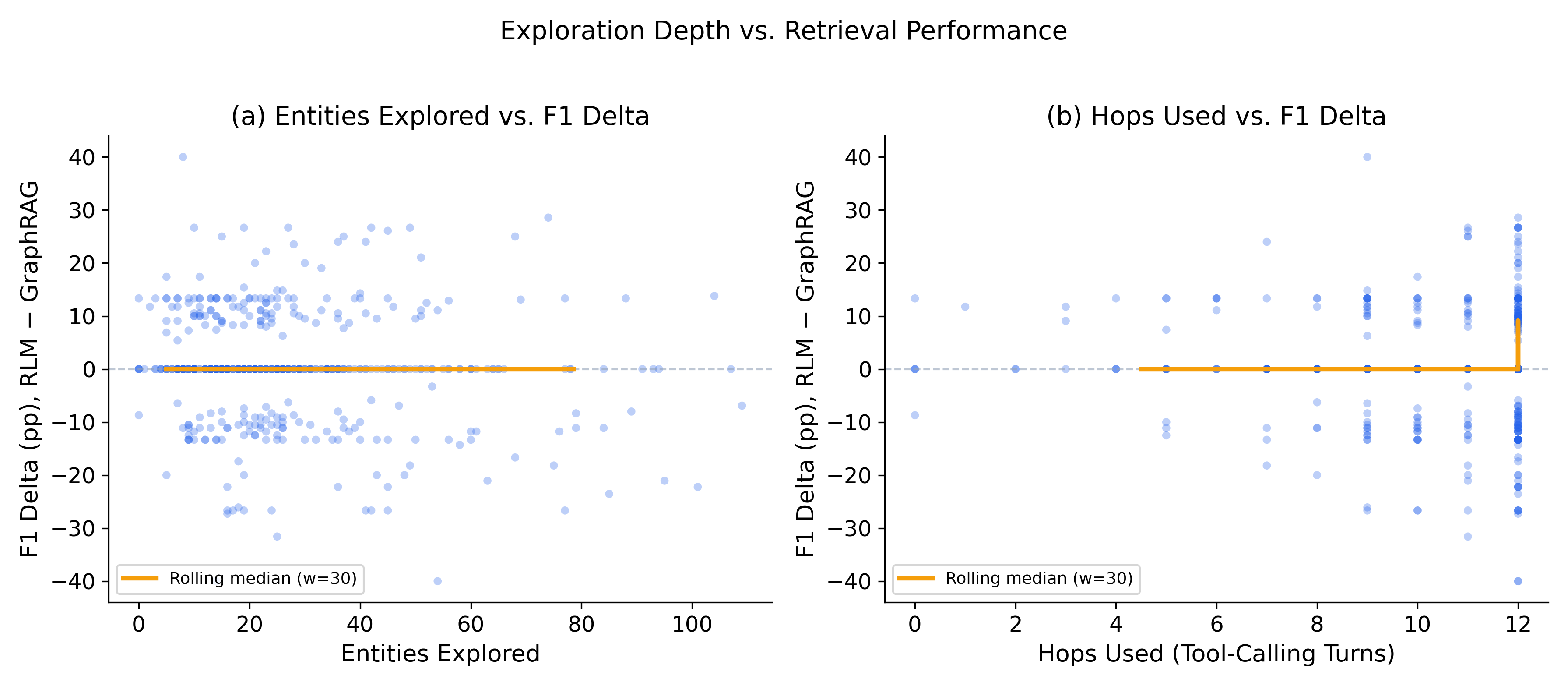}}
\caption{Exploration depth (entities visited) vs retrieval F1. Moderate
exploration (\textasciitilde20--30 entities) yields the best results.}
\end{figure}

\subsubsection{5.6 When Does RLM-on-KG
Excel?}\label{when-does-rlm-on-kg-excel}

The overall F1 delta (+0.16 pp) masks a more nuanced story. To
understand \emph{which classes of questions} benefit most from KG-guided
exploration, we cross-tabulate RLM's win rate by question type and
\textbf{evidence scatter} --- the number of gold evidence chunks a
question requires.

\textbf{Win rate by question type $\times$ evidence scatter.}

{\def\LTcaptype{none} % do not increment counter
\begin{longtable}[]{@{}
  >{\raggedright\arraybackslash}p{(\linewidth - 6\tabcolsep) * \real{0.1622}}
  >{\raggedright\arraybackslash}p{(\linewidth - 6\tabcolsep) * \real{0.2703}}
  >{\raggedright\arraybackslash}p{(\linewidth - 6\tabcolsep) * \real{0.2973}}
  >{\raggedright\arraybackslash}p{(\linewidth - 6\tabcolsep) * \real{0.2703}}@{}}
\toprule\noalign{}
\begin{minipage}[b]{\linewidth}\raggedright
Type
\end{minipage} & \begin{minipage}[b]{\linewidth}\raggedright
Gold 1--5
\end{minipage} & \begin{minipage}[b]{\linewidth}\raggedright
Gold 6--10
\end{minipage} & \begin{minipage}[b]{\linewidth}\raggedright
Gold 11+
\end{minipage} \\
\midrule\noalign{}
\endhead
\bottomrule\noalign{}
\endlastfoot
Fact Retrieval & 57\% (39W/136T/29L) & 57\% (13W/22T/10L) &
\textbf{100\%} (3W/3T/0L) \\
Complex Reasoning & 33\% (1W/8T/2L) & 44\% (24W/57T/31L) & \textbf{69\%}
(18W/7T/8L) \\
Contextual Summ. & n\textless5 & 48\% (10W/15T/11L) & 46\%
(16W/18T/19L) \\
\end{longtable}
}

Win rate = RLM wins $\div$ (RLM wins + GraphRAG wins), excluding ties.

Two patterns emerge:

\begin{enumerate}
\def\labelenumi{\arabic{enumi}.}
\item
  \textbf{Evidence scatter is the strongest predictor of RLM advantage.}
  When the gold evidence is distributed across 11+ chunks, RLM-on-KG's
  win rate rises to 56\% overall (71 non-ties), with a mean F1 delta of
  +1.38 pp.~For Fact Retrieval with 11+ gold chunks, the win rate
  reaches 100\% (though n=6). The mechanism is clear: multi-hop KG
  traversal follows entity links to discover chunks that are
  semantically distant from the query but structurally connected ---
  precisely the chunks that vector search misses.
\item
  \textbf{The question type modulates the effect.} Fact Retrieval
  benefits most because fact-bearing chunks are often entity-specific
  (e.g., a character's backstory mentioned in a different chapter).
  Complex Reasoning benefits when evidence is scattered (69\% win rate
  at 11+ chunks) but suffers when evidence is concentrated (33\% at 1--5
  chunks), where over-exploration adds noise. The n=6 result for
  high-scatter Fact Retrieval (100\% win rate) is directional but should
  be interpreted with caution due to the very small sample size.
\end{enumerate}

\textbf{Qualitative characterization of RLM's sweet spot.}

{\def\LTcaptype{none} % do not increment counter
\begin{longtable}[]{@{}
  >{\raggedright\arraybackslash}p{(\linewidth - 4\tabcolsep) * \real{0.3409}}
  >{\raggedright\arraybackslash}p{(\linewidth - 4\tabcolsep) * \real{0.2727}}
  >{\raggedright\arraybackslash}p{(\linewidth - 4\tabcolsep) * \real{0.3864}}@{}}
\toprule\noalign{}
\begin{minipage}[b]{\linewidth}\raggedright
Characteristic
\end{minipage} & \begin{minipage}[b]{\linewidth}\raggedright
RLM excels
\end{minipage} & \begin{minipage}[b]{\linewidth}\raggedright
GraphRAG excels
\end{minipage} \\
\midrule\noalign{}
\endhead
\bottomrule\noalign{}
\endlastfoot
Evidence distribution & Scattered (11+ chunks, multiple chapters) &
Concentrated (1--5 chunks, single passage) \\
Required traversal & Multi-hop entity chains & Single-hop or direct
similarity \\
Question structure & ``What happens across\ldots{}'', ``Trace
the\ldots{}'' & ``What does X say about\ldots{}'' \\
Entity involvement & Multiple entities co-mentioned & Single entity or
no entity needed \\
RLM recall advantage & +1.2 pp (11+ chunks) & $-$0.7 pp (6--10 chunks) \\
\end{longtable}
}

When RLM wins, the recall gap is the driver: RLM achieves 61.9\% recall
vs GraphRAG's 45.8\% on winning questions, indicating that KG traversal
discovers relevant chunks that similarity search cannot reach. When RLM
loses, the pattern reverses --- over-exploration introduces irrelevant
chunks that dilute precision without improving recall.

\begin{figure}
\centering
\pandocbounded{\includegraphics[keepaspectratio,alt={Win rate heatmap by question type $\times$ evidence scatter. Darker cells indicate stronger RLM advantage. The combination of Fact Retrieval + 11+ gold chunks shows the highest win rate.}]{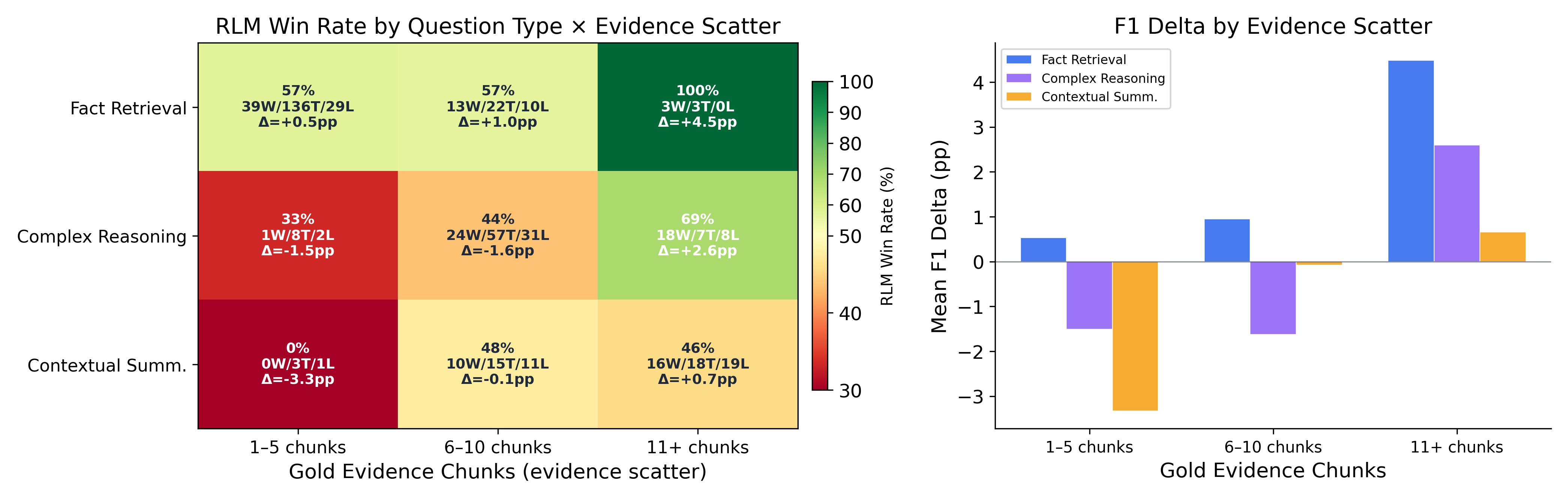}}
\caption{Win rate heatmap by question type $\times$ evidence scatter. Darker
cells indicate stronger RLM advantage. The combination of Fact Retrieval
+ 11+ gold chunks shows the highest win rate.}
\end{figure}

\begin{figure}
\centering
\pandocbounded{\includegraphics[keepaspectratio,alt={Win rate by evidence scatter. As the required number of gold chunks increases, RLM-on-KG's probability of beating GraphRAG rises significantly, reaching 56\% for 11+ chunks.}]{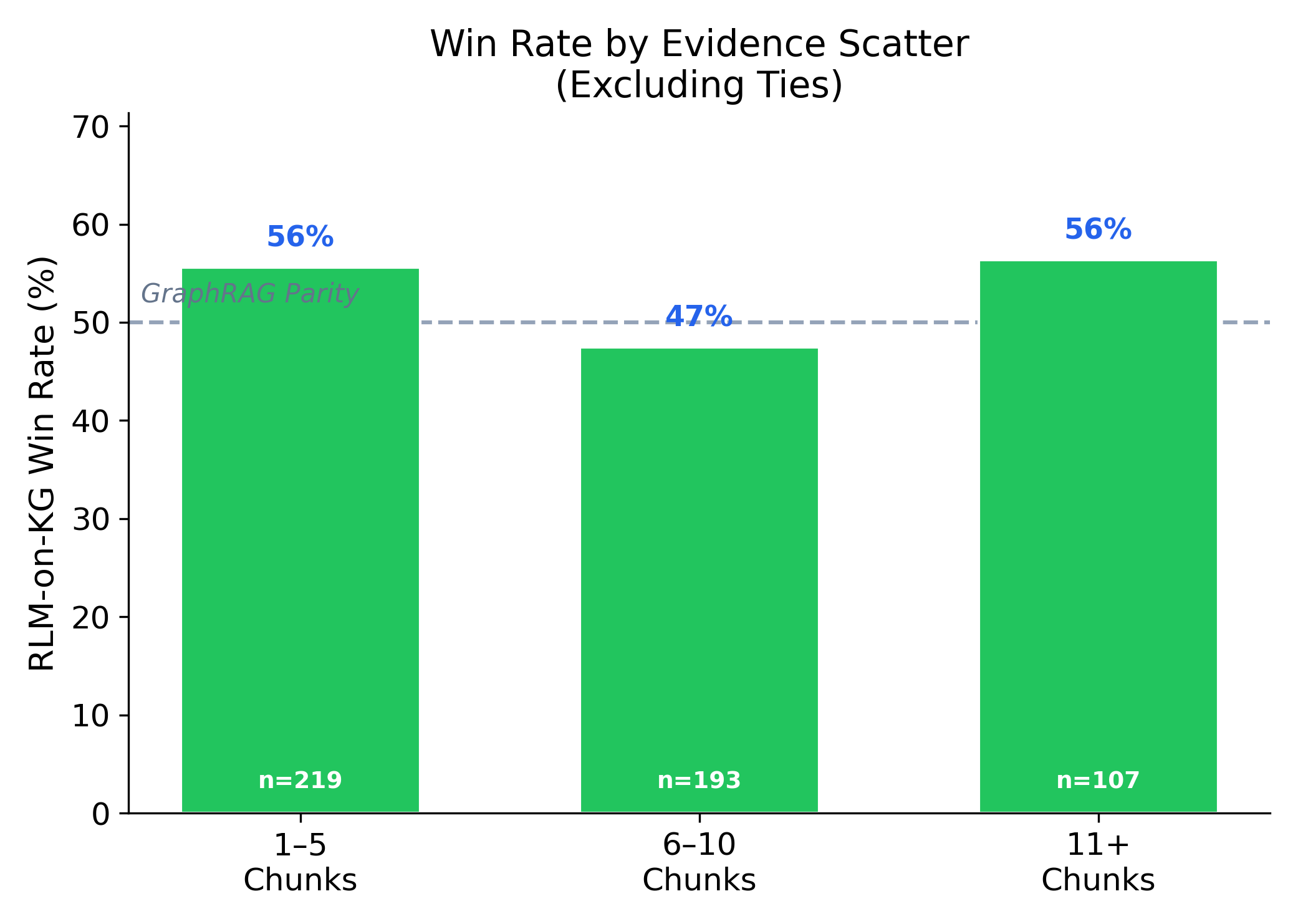}}
\caption{Win rate by evidence scatter. As the required number of gold
chunks increases, RLM-on-KG's probability of beating GraphRAG rises
significantly, reaching 56\% for 11+ chunks.}
\end{figure}

\begin{figure}
\centering
\pandocbounded{\includegraphics[keepaspectratio,alt={Recall advantage vs evidence scatter. RLM-on-KG's ability to discover missing chunks outpaces GraphRAG specifically on highly scattered questions.}]{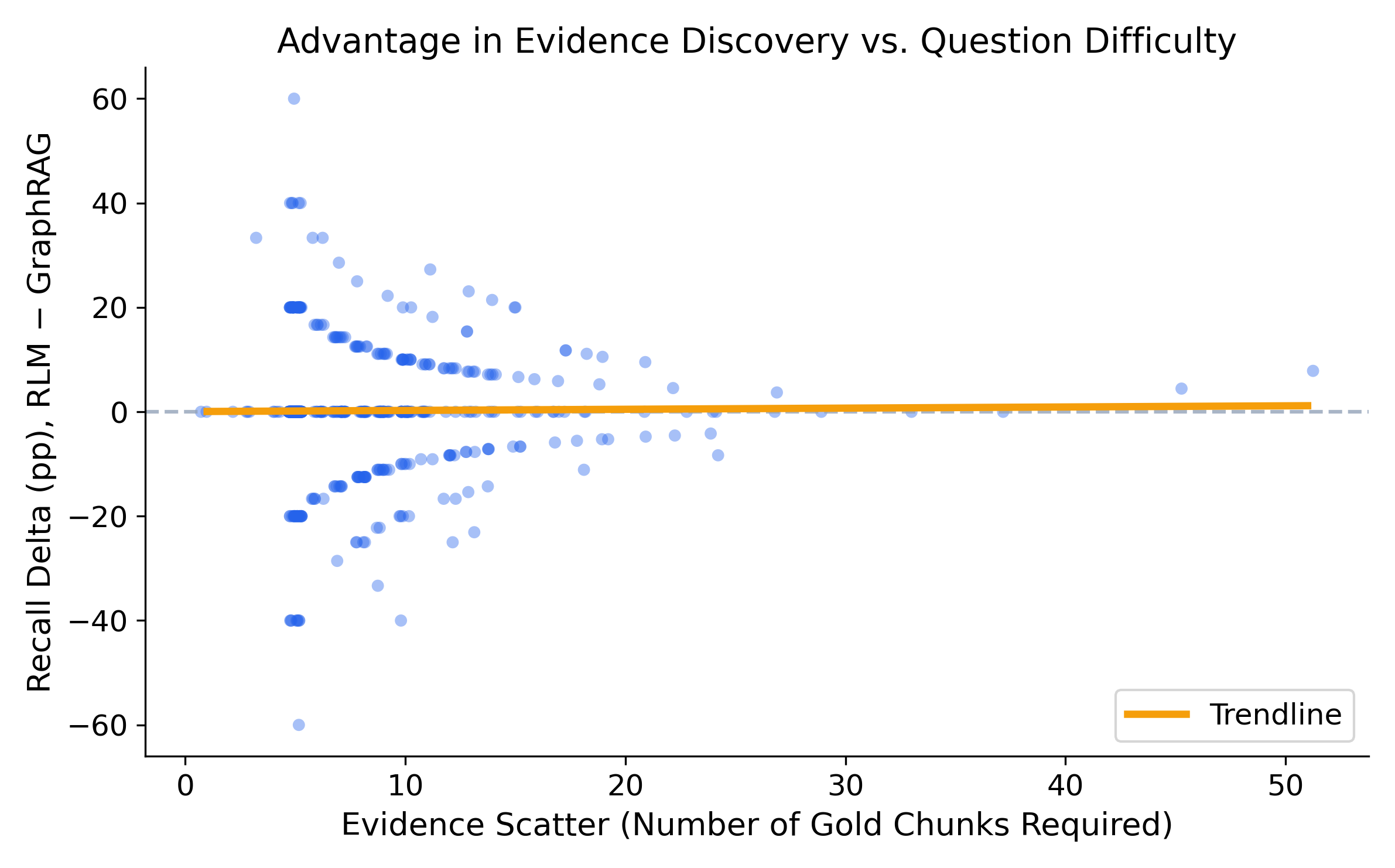}}
\caption{Recall advantage vs evidence scatter. RLM-on-KG's ability to
discover missing chunks outpaces GraphRAG specifically on highly
scattered questions.}
\end{figure}

\subsubsection{5.7 Exploration Trace
Examples}\label{exploration-trace-examples}

To illustrate the behavioral difference between LLM-driven and heuristic
traversal, we present two representative cases from the 519-question
evaluation: one where the LLM controller wins decisively, and one where
it fails.

\textbf{Success case: LLM discovers distant evidence (Complex Reasoning,
13 gold chunks).} The heuristic baseline explored 27 entities in 2 hops
but saturated at hop 1, achieving F1 = 0.348. The LLM controller
explored 32 entities across 12 hops (18 tool calls, 1 sub-query),
achieving F1 = 0.609 (+26.1 pp). The mechanism: after the initial entity
expansion covered nearby evidence, the LLM pursued indirect entity
connections (characters' associates, co-mentioned locations) that the
heuristic's breadth-first strategy did not prioritize. This additional
exploration surfaced 8 gold chunks that the heuristic missed --- chunks
semantically distant from the query but structurally connected through
the mention graph. GraphRAG achieved F1 = 0.522 on this question,
showing that even its 1-hop expansion missed the most distant evidence.

\textbf{Failure case: LLM over-explores and dilutes (Complex Reasoning,
10 gold chunks).} The heuristic baseline explored 26 entities in 2 hops
and achieved F1 = 0.800. The LLM controller explored 54 entities across
12 hops (34 tool calls, 1 sub-query) and achieved F1 = 0.400 ($-$40.0 pp).
The failure mode: the LLM followed tangentially related entity paths
(exploring peripheral characters, cross-chapter references) that
introduced non-gold chunks into the candidate pool, diluting precision
without improving recall. The heuristic's early saturation was actually
advantageous --- the correct evidence was concentrated near the seed
entities, and broader exploration only added noise. GraphRAG achieved F1
= 0.800, confirming that 1-hop expansion was sufficient.

\textbf{Pattern summary.} These cases illustrate the conditional
advantage thesis: the LLM controller wins when evidence is scattered
across entity neighborhoods that require adaptive, multi-step navigation
to reach. It loses when evidence is concentrated and the exploration
budget is better spent on precision than breadth. The heuristic's 2-hop
/ early saturation behavior is a feature, not a bug, for concentrated
evidence; the LLM's 12-hop / deep exploration is essential for scattered
evidence.

\subsubsection{5.8 Cross-Model Validation}\label{cross-model-validation}

\paragraph{5.8.1 Claude Haiku 4.5 at Full Scale (519
questions)}\label{claude-haiku-4.5-at-full-scale-519-questions}

To test whether the LLM controller advantage transfers across model
families, we run the complete 519-question evaluation with Claude Haiku
4.5 (Anthropic, via Vertex AI) as the controller. The same shared KG,
tools, system prompt, and evaluation pipeline are used as in the primary
Gemini results (Section 5.4--5.5).

\textbf{Results.}

{\def\LTcaptype{none} % do not increment counter
\begin{longtable}[]{@{}lccc@{}}
\toprule\noalign{}
System & Precision & Recall & F1 \\
\midrule\noalign{}
\endhead
\bottomrule\noalign{}
\endlastfoot
Vector-only (top-20) & --- & --- & 38.4\% \\
GraphRAG-local variant & --- & --- & 44.8\% \\
Heuristic RLM & --- & --- & 42.8\% \\
\textbf{RLM-on-KG (Claude)} & --- & --- & \textbf{47.2\%} \\
\end{longtable}
}

Head-to-head (Claude LLM vs Heuristic): \textbf{205 wins, 239 ties, 75
losses} (win rate 73.2\% among decisive questions).

Head-to-head (Claude LLM vs GraphRAG): \textbf{170 wins, 259 ties, 90
losses} (win rate 65.4\% among decisive questions).

\textbf{Statistical context.} Bootstrap 95\% confidence intervals
(B=10,000):

{\def\LTcaptype{none} % do not increment counter
\begin{longtable}[]{@{}
  >{\raggedright\arraybackslash}p{(\linewidth - 6\tabcolsep) * \real{0.3158}}
  >{\centering\arraybackslash}p{(\linewidth - 6\tabcolsep) * \real{0.2368}}
  >{\centering\arraybackslash}p{(\linewidth - 6\tabcolsep) * \real{0.2105}}
  >{\centering\arraybackslash}p{(\linewidth - 6\tabcolsep) * \real{0.2368}}@{}}
\toprule\noalign{}
\begin{minipage}[b]{\linewidth}\raggedright
Comparison
\end{minipage} & \begin{minipage}[b]{\linewidth}\centering
Mean $\Delta$F1
\end{minipage} & \begin{minipage}[b]{\linewidth}\centering
95\% CI
\end{minipage} & \begin{minipage}[b]{\linewidth}\centering
p-value
\end{minipage} \\
\midrule\noalign{}
\endhead
\bottomrule\noalign{}
\endlastfoot
\textbf{Claude LLM $-$ Heuristic} & \textbf{+4.37 pp} & \textbf{{[}+3.33,
+5.44{]}} & \textbf{\textless{} 0.001} \\
\textbf{Claude LLM $-$ GraphRAG} & \textbf{+2.42 pp} & \textbf{{[}+1.51,
+3.33{]}} & \textbf{\textless{} 0.001} \\
\end{longtable}
}

\textbf{This is the paper's strongest result.} With Claude Haiku as
controller, RLM-on-KG not only beats the heuristic baseline (+4.37 pp, p
\textless{} 0.001) but also \textbf{statistically significantly
outperforms GraphRAG-local} (+2.42 pp, p \textless{} 0.001). The Gemini
primary results showed a non-significant advantage over GraphRAG (+0.16
pp, p=0.36); with Claude, the advantage is both larger and statistically
significant, confirming that the LLM controller's value scales with
tool-calling capability.

\paragraph{5.8.2 Cross-Model Comparison (100 questions $\times$ 3
controllers)}\label{cross-model-comparison-100-questions-3-controllers}

To map the relationship between controller capability and retrieval
performance, we compare three model families on 100 questions:

{\def\LTcaptype{none} % do not increment counter
\begin{longtable}[]{@{}lcccc@{}}
\toprule\noalign{}
Controller & Heuristic F1 & LLM F1 & $\Delta$ (LLM$-$Heur) & Latency \\
\midrule\noalign{}
\endhead
\bottomrule\noalign{}
\endlastfoot
\textbf{Claude Haiku 4.5} & 0.376 & \textbf{0.412} & \textbf{+3.60 pp} &
\dag{} \\
Gemini 2.5 Flash Lite & 0.371 & 0.380 & +0.84 pp & 68 s \\
Gemma 4 E2B (local) & 0.350 & 0.343 & $-$0.78 pp & 237 s \\
\end{longtable}
}

\dag{} \emph{Claude latency was not reliably recorded for this run (logging
error); the MuSiQue evaluation (Section 5.9) recorded \textasciitilde20
s per question on smaller per-question KGs, suggesting Novel latency
would be higher.}

\textbf{Bootstrap 95\% CIs (n=100, B=10,000):}

{\def\LTcaptype{none} % do not increment counter
\begin{longtable}[]{@{}
  >{\raggedright\arraybackslash}p{(\linewidth - 6\tabcolsep) * \real{0.2449}}
  >{\centering\arraybackslash}p{(\linewidth - 6\tabcolsep) * \real{0.4082}}
  >{\centering\arraybackslash}p{(\linewidth - 6\tabcolsep) * \real{0.1633}}
  >{\centering\arraybackslash}p{(\linewidth - 6\tabcolsep) * \real{0.1837}}@{}}
\toprule\noalign{}
\begin{minipage}[b]{\linewidth}\raggedright
Controller
\end{minipage} & \begin{minipage}[b]{\linewidth}\centering
Mean $\Delta$ (LLM$-$Heur)
\end{minipage} & \begin{minipage}[b]{\linewidth}\centering
95\% CI
\end{minipage} & \begin{minipage}[b]{\linewidth}\centering
p-value
\end{minipage} \\
\midrule\noalign{}
\endhead
\bottomrule\noalign{}
\endlastfoot
\textbf{Claude Haiku 4.5} & \textbf{+3.60 pp} & \textbf{{[}+2.16,
+5.18{]}} & \textbf{\textless{} 0.001} \\
Gemini 2.5 Flash Lite & +0.84 pp & {[}$-$1.75, +3.34{]} & 0.52 \\
Gemma 4 E2B (local) & $-$0.78 pp & {[}$-$2.57, +0.85{]} & 0.45 \\
\end{longtable}
}

\textbf{Key findings.}

\begin{enumerate}
\def\labelenumi{\arabic{enumi}.}
\item
  \textbf{The advantage scales monotonically with tool-calling
  capability}: Claude (+3.60 pp) \textgreater{} Gemini (+0.84 pp)
  \textgreater{} Gemma E2B ($-$0.78 pp). This ordering is consistent
  across both evaluation sets (see Section 5.9).
\item
  \textbf{Gemma 4 E2B achieves an 83\% behavioral tie rate} with the
  heuristic. Of the 100 questions, 53 did not complete (the model did
  not emit a final response within the per-question timeout); these were
  assigned the heuristic's F1 and counted as ties in the bootstrap,
  preserving n=100. Among the 47 completed questions: 3 wins, 39 ties, 5
  losses. The high non-completion rate (53\%) is consistent with early
  Gemma 4 tool-calling limitations under local quantization, though we
  cannot rule out question-difficulty bias in the non-completed subset.
\item
  \textbf{Latency varies substantially} across controllers: Gemma 4 E2B
  averages 237 s per question; Gemini 2.5 Flash Lite averages 68 s.
  Claude latency was not reliably recorded for this run (see footnote
  above). The spread across reliably measured controllers is roughly
  3.5$\times$.
\end{enumerate}

\textbf{Cost-normalized analysis.} Per-question cost (estimated at
\textasciitilde50K tokens/question, \textasciitilde80\% input):

{\def\LTcaptype{none} % do not increment counter
\begin{longtable}[]{@{}
  >{\raggedright\arraybackslash}p{(\linewidth - 8\tabcolsep) * \real{0.1739}}
  >{\centering\arraybackslash}p{(\linewidth - 8\tabcolsep) * \real{0.2609}}
  >{\centering\arraybackslash}p{(\linewidth - 8\tabcolsep) * \real{0.1739}}
  >{\centering\arraybackslash}p{(\linewidth - 8\tabcolsep) * \real{0.1304}}
  >{\centering\arraybackslash}p{(\linewidth - 8\tabcolsep) * \real{0.2609}}@{}}
\toprule\noalign{}
\begin{minipage}[b]{\linewidth}\raggedright
Controller
\end{minipage} & \begin{minipage}[b]{\linewidth}\centering
Delta (LLM-Heur)
\end{minipage} & \begin{minipage}[b]{\linewidth}\centering
\$/question
\end{minipage} & \begin{minipage}[b]{\linewidth}\centering
Latency
\end{minipage} & \begin{minipage}[b]{\linewidth}\centering
Delta per dollar
\end{minipage} \\
\midrule\noalign{}
\endhead
\bottomrule\noalign{}
\endlastfoot
Claude Haiku 4.5 & +3.60 pp & \(0.072 | n/a | ~+50 pp/\) & & \\
Gemini 2.5 Flash Lite & +0.84 pp & \(0.006 | 68 s | ~+140 pp/\) & & \\
Gemma 4 E2B (local) & -0.78 pp & \$0.000 & 237 s & n/a \\
\end{longtable}
}

For context, full GraphRAG offline indexing (LLM entity extraction +
community summaries on 20 novels) costs an estimated \$15 as a one-time
cost, amortized across all subsequent queries on a stable corpus.
RLM-on-KG with Gemini on 519 questions costs \$3.11 total in query-time
tokens --- a favorable \textbf{deployment tradeoff for evolving corpora}
where the offline index must be refreshed frequently and its cost cannot
be amortized. For stable corpora with many repeated queries,
GraphRAG-local remains more cost-efficient.

\textbf{Interpretation.} The cross-model results reveal that the LLM
controller advantage is not a property of any specific model but of
\textbf{tool-calling sophistication}. The monotonic ordering (Claude
\textgreater{} Gemini \textgreater{} Gemma) and the Gemma behavioral tie
rate establish a clear spectrum: models with strong function-calling and
multi-turn reasoning exhibit adaptive exploration; models with weaker
tool-calling default to heuristic-like patterns. This behavioral gap
motivates a distillation approach (Section 6).

\subsubsection{5.9 Cross-Scale Robustness Check (MuSiQue, 50
questions)}\label{cross-scale-robustness-check-musique-50-questions}

To test whether the LLM controller advantage scales to a different graph
structure and domain, we evaluate on MuSiQue (Trivedi et al., 2022), a
multi-hop decomposable QA benchmark. Each question's supporting and
distractor paragraphs are used to construct a \textbf{per-question KG
(\textasciitilde20 chunks)} rather than the shared corpus KG used for
Novel (\textasciitilde1,000+ chunks). This structural difference is key:
smaller per-question KGs have fewer hub entities and lower co-mention
density, which limits the LLM controller's structural-navigation
advantage.

\textbf{Results.}

{\def\LTcaptype{none} % do not increment counter
\begin{longtable}[]{@{}
  >{\raggedright\arraybackslash}p{(\linewidth - 10\tabcolsep) * \real{0.1690}}
  >{\centering\arraybackslash}p{(\linewidth - 10\tabcolsep) * \real{0.1831}}
  >{\centering\arraybackslash}p{(\linewidth - 10\tabcolsep) * \real{0.1972}}
  >{\centering\arraybackslash}p{(\linewidth - 10\tabcolsep) * \real{0.1127}}
  >{\centering\arraybackslash}p{(\linewidth - 10\tabcolsep) * \real{0.2113}}
  >{\centering\arraybackslash}p{(\linewidth - 10\tabcolsep) * \real{0.1268}}@{}}
\toprule\noalign{}
\begin{minipage}[b]{\linewidth}\raggedright
Controller
\end{minipage} & \begin{minipage}[b]{\linewidth}\centering
GraphRAG F1
\end{minipage} & \begin{minipage}[b]{\linewidth}\centering
Heuristic F1
\end{minipage} & \begin{minipage}[b]{\linewidth}\centering
LLM F1
\end{minipage} & \begin{minipage}[b]{\linewidth}\centering
$\Delta$ (LLM$-$Heur)
\end{minipage} & \begin{minipage}[b]{\linewidth}\centering
Latency
\end{minipage} \\
\midrule\noalign{}
\endhead
\bottomrule\noalign{}
\endlastfoot
\textbf{Claude Haiku 4.5} & 0.355 & 0.311 & \textbf{0.327} &
\textbf{+1.67 pp} & 20 s \\
Gemini 2.5 Flash Lite & 0.356 & 0.311 & 0.324 & +1.33 pp & 27 s \\
\end{longtable}
}

\textbf{Key findings.}

\begin{enumerate}
\def\labelenumi{\arabic{enumi}.}
\item
  \textbf{Both API models show a consistent positive delta} over
  heuristic traversal on MuSiQue, confirming the LLM controller
  advantage is not unique to the Novel benchmark.
\item
  \textbf{Claude again outperforms Gemini} (+1.67 pp vs +1.33 pp),
  consistent with the Novel results.
\item
  The \textbf{absolute F1 values are lower} than Novel
  (\textasciitilde0.33 vs \textasciitilde0.40), which is expected:
  MuSiQue per-question KGs contain only \textasciitilde20 chunks with a
  high distractor ratio, making retrieval harder.
\item
  \textbf{Cross-scale attenuation consistent with theory.} Claude's
  MuSiQue delta (+1.67 pp) is 62\% smaller than its Novel delta (+4.37
  pp). This is exactly what the paper's theory predicts: on smaller,
  lower-hub graphs with fewer co-mention chains, the LLM controller's
  structural- navigation advantage has less room to operate. We
  interpret this as cross-scale attenuation, not cross-benchmark
  validation of identical gains.
\item
  \textbf{LLM RLM trails GraphRAG-local on MuSiQue} ($-$2.8 pp for Claude,
  $-$3.2 pp for Gemini). This is a scope limitation of the main claim: the
  +2.42 pp vs GraphRAG advantage established for Claude at n=519
  (Section 5.8.1) does not transfer to 20-chunk per-question KGs. The
  GraphRAG-local pipeline's 1-hop entity expansion is better matched to
  the small, distractor-dominated graph structure of MuSiQue, while the
  LLM controller's deeper exploration introduces more noise when the
  graph has few meaningful hops to exploit.
\end{enumerate}

\begin{center}\rule{0.5\linewidth}{0.5pt}\end{center}

\subsection{6. Discussion}\label{discussion}

\subsubsection{When RLM-on-KG wins}\label{when-rlm-on-kg-wins}

RLM-on-KG is most effective for questions that require reaching chunks
not directly similar to the query but connected through entity
relationships. At full scale, the largest per-type gain is on
\textbf{Fact Retrieval} (+0.7 pp, 55 wins vs 39 losses), where
entity-driven traversal surfaces chunks containing specific facts
scattered across the corpus. The largest single-question improvements
(up to +40 pp F1) occur when the gold evidence is distributed across
chunks mentioning different entities involved in the same event or
relationship.

This ability to traverse structural links to find distant evidence
aligns with findings in other domains where graph reasoning surfaces
less obvious connections. For example, Buehler (2024) used path sampling
over a knowledge graph to reveal structural parallels between biological
materials and Beethoven's 9th Symphony. Similarly, our results show that
RLM-on-KG excels precisely when the required evidence is highly
scattered (11+ chunks apart) and unattainable by vector search.

This makes RLM-on-KG particularly suited to: - Multi-hop questions
requiring evidence from entity intersections - Fact retrieval across
narratively distant text passages - Evolving corpora where reindexing
costs make offline pipelines impractical - Applications requiring
provenance: each retrieved chunk links to its source entity and document
via RDF identifiers, supporting citation and auditability

\subsubsection{When GraphRAG wins}\label{when-graphrag-wins}

The GraphRAG-local variant excels in low-latency settings. Its heuristic
retrieval completes in under 1 second with zero LLM calls, making it
suitable for interactive applications. It also benefits from amortized
cost: once the vector index is built, repeated queries on a stable
corpus are essentially free. For simple single-hop questions where the
answer appears in a chunk directly similar to the query, the additional
exploration of RLM-on-KG provides no benefit and adds unnecessary
latency and cost.

\subsubsection{Hybrid opportunity}\label{hybrid-opportunity}

The two approaches are complementary: the balanced head-to-head record
(127 wins, 277 ties, 115 losses vs GraphRAG) confirms that neither
system dominates overall. The more informative comparison is LLM
vs.~Heuristic RLM (176W/241T/102L, p \textless{} 0.0001), which shows
that the LLM controller consistently adds value. The advantage is
conditional: for questions whose gold evidence is scattered across 11+
chunks, the LLM controller's win rate over heuristic traversal is 62\%.
If GraphRAG is a static expansion graph, RLM-on-KG is a \textbf{control
policy over that graph} --- one that activates selectively when the
retrieval problem demands structural navigation. A practical deployment
could use the GraphRAG-local variant as a fast first pass, then invoke
RLM-on-KG when scatter indicators suggest the answer requires evidence
from multiple entity neighborhoods.

\subsubsection{Discovery vs.~ranking}\label{discovery-vs.-ranking}

The most important architectural lesson from our iterative optimization
(V1--V7) is the separation between \textbf{candidate discovery} and
\textbf{ranking}. Early iterations attempted to use entity-derived
scores, co-mention boosts, and MMR diversity penalties to improve
ranking. These refinements either had no effect or \emph{degraded}
performance. The breakthrough came from simplifying: let the LLM's
multi-hop mention-graph traversal discover a broad pool of candidate
chunks, then rank them with the same pure vector similarity that
GraphRAG uses. This insight --- that LLM exploration adds value through
discovery, not judgment --- suggests that future work should focus on
expanding the quality of the candidate pool (e.g., via cross-encoder
re-ranking) rather than complex scoring formulas.

\textbf{One-liner takeaway:} Separate discovery from ranking; let the
LLM explore but let the vectors decide.

\subsubsection{Selective escalation}\label{selective-escalation}

The controller continuum (vector-only $\rightarrow$ GraphRAG-local $\rightarrow$ heuristic RLM $\rightarrow$
LLM RLM) suggests a practical deployment strategy: \textbf{selective
escalation}. Use the cheapest sufficient controller for each question.
GraphRAG-local handles \textasciitilde53\% of questions as well as the
LLM controller (the tie rate). For the remaining 47\%, evidence scatter
indicators (e.g., seed entity count, initial recall gap) could trigger
escalation to LLM-guided traversal. A learned routing policy trained on
exploration traces could automate this decision, reducing average cost
to near GraphRAG-local levels while preserving LLM-level quality on hard
questions.

\subsubsection{Exploration traces as
supervision}\label{exploration-traces-as-supervision}

The cross-model results reveal that the gap between Gemma 4 E2B (83\%
behavioral tie rate) and Claude (+4.37 pp) is not architectural but
behavioral: both use the same tools, but Claude makes adaptive
exploration decisions while Gemma defaults to shallow patterns. Each
question produces a complete (state, action, reward) trajectory that can
serve as training signal. Concretely:

\begin{itemize}
\tightlist
\item
  \textbf{Training data}: Claude exploration traces on MuSiQue and Novel
  provide \textasciitilde519 trajectories of multi-turn tool-calling
  sequences, with per-question F1 as reward signal.
\item
  \textbf{Target model}: Gemma 4 E2B or E4B, fine-tuned with TRL
  (Transformer Reinforcement Learning) to close the behavioral gap.
\item
  \textbf{Routing policy}: The same traces can train a lightweight
  classifier to predict when LLM exploration will outperform heuristic
  traversal, enabling selective escalation without running the full LLM
  loop.
\end{itemize}

This positions the current work not just as a retrieval system but as a
\textbf{data collection pipeline} for training smaller, deployable KG
navigators. Effective use of these traces also requires that target
models manage their context window across multi-turn exploration
sequences; techniques for teaching LLMs to manage their own context
(Kontonis et al., 2026) are directly applicable to this training regime.

\subsubsection{Efficiency strategies}\label{efficiency-strategies}

The current per-question cost (\textasciitilde50K tokens, 2--5 minutes)
is a significant limitation but not a fundamental barrier. Several
reduction strategies are available:

\begin{itemize}
\tightlist
\item
  \textbf{Caching}: \texttt{entity\_search} and
  \texttt{get\_chunks\_for\_entity} results can be cached across
  questions sharing the same seed entities, reducing redundant KG
  lookups.
\item
  \textbf{Degree-bounded expansion}: Capping \texttt{expand\_neighbors}
  by node degree and filtering by edge relevance prevents exploration
  drift into densely connected but uninformative entity neighborhoods.
\item
  \textbf{Early stopping on score saturation}: Monitoring composite
  score growth across turns and terminating when marginal improvements
  fall below a threshold can halve the average turn count with minimal
  F1 impact.
\item
  \textbf{Model tiering}: Using a smaller, faster model for navigation
  and tool selection, and a stronger model only for final answer
  generation, can reduce latency and cost by an estimated 3--5$\times$.
\end{itemize}

\subsubsection{Limitations}\label{limitations}

\begin{itemize}
\tightlist
\item
  \textbf{Latency}: RLM-on-KG requires 2 to 5 minutes per question (10
  to 25 LLM calls), compared to sub-second retrieval for the
  GraphRAG-local variant. This makes it unsuitable for real-time
  interactive applications in its current form.
\item
  \textbf{Cost}: at approximately 50K tokens per question, RLM-on-KG is
  roughly 25$\times$ more expensive than the GraphRAG-local variant on a
  per-query basis.
\item
  \textbf{Exploration drift}: the LLM may follow entity paths that lead
  away from relevant evidence. Our stall detection mitigates this but
  does not eliminate it. Complex Reasoning questions show a slight
  negative delta ($-$0.7 pp vs GraphRAG at full scale), likely because
  broad exploration introduces noise for questions with concentrated
  evidence.
\item
  \textbf{Gold label noise}: All gold evidence maps through semantic
  similarity (threshold = 0.25) using the same embedding model as
  retrieval, introducing noise at low similarity scores and a
  shared-representation confound. While head-to-head comparisons remain
  valid, absolute F1 values should be interpreted with caution (Section
  4.3).
\item
  \textbf{Statistical significance}: On 519 questions, the overall RLM
  vs GraphRAG delta (+0.16 pp) is not statistically significant
  (bootstrap p=0.36). The LLM vs Heuristic delta (+2.47 pp) \emph{is}
  statistically significant (p \textless{} 0.0001), supporting the core
  architectural claim.
\item
  \textbf{Memorization risk}: The GraphRAG-Bench Novel corpus consists
  of public-domain fiction texts that are likely present in the training
  data of Gemini 2.0 Flash. Since RLM-on-KG uses this model at query
  time for navigation decisions, parametric knowledge could influence
  tool selection in ways that do not generalize to novel corpora.
  Importantly, the LLM cannot fabricate evidence --- it can only collect
  chunks that exist in the KG --- and final ranking uses pure vector
  similarity, not LLM judgment. Nevertheless, evaluation on
  non-public-domain corpora would strengthen the external validity of
  these results.
\item
  \textbf{Controlled GraphRAG comparison}: Our GraphRAG-local variant
  intentionally removes community detection, community summaries, and
  LLM-based entity extraction to isolate retrieval logic. This is not a
  comparison with the full GraphRAG system; performance against the
  complete pipeline may differ.
\item
  \textbf{Mention graph only}: Our graph captures entity mentions, not
  typed semantic relations. Systems with richer relation extraction may
  achieve better performance on questions requiring relation-level
  reasoning.
\item
  \textbf{Benchmark scope}: The primary results (Sections 5.1--5.6) are
  on a single benchmark (GraphRAG-Bench Novel) focusing on fiction
  texts. Cross-scale validation on MuSiQue (Section 5.9) shows that the
  LLM-over-heuristic delta transfers but attenuates (\textasciitilde62\%
  smaller on 20-chunk per-question KGs); evaluation on scientific,
  medical, and legal domains remains future work.
\item
  \textbf{Cross-model coverage}: We evaluate three model families
  (Claude, Gemini, Gemma 4) in Section 5.8. The LLM controller advantage
  holds for API-served models with strong tool-calling capabilities but
  not for local models with weaker function-calling behavior. Coverage
  of additional model families (e.g., Llama, Mistral) remains future
  work.
\item
  \textbf{Cross-model sample size}: The cross-model experiments (Section
  5.8) use n=100 questions (Novel) and n=50 (MuSiQue), smaller than the
  primary n=519 evaluation. Bootstrap confidence intervals are not
  computed for these subsets.
\item
  \textbf{Open-source model deployment boundary}: For current
  locally-runnable open-source models (Gemma 4 E2B in our evaluation),
  LLM control is net negative ($-$0.78 pp, 53\% non-completion rate). This
  is a deployment boundary, not only a distillation opportunity:
  practitioners deploying on-device KG navigation should default to the
  heuristic baseline until model tool-calling capability matures or
  distilled navigators are available.
\end{itemize}

\begin{center}\rule{0.5\linewidth}{0.5pt}\end{center}

\subsection{7. Conclusion}\label{conclusion}

We have studied when an LLM controller outperforms rule-based heuristic
traversal for knowledge graph exploration. The answer is: \textbf{when
evidence is scattered and the model has strong tool-calling capability}.

On GraphRAG-Bench Novel (519 questions), LLM-driven traversal achieves
+2.47 pp F1 over a heuristic baseline using the same graph and tools (p
\textless{} 0.0001). The gain is largest at moderate scatter (+3.21 pp
at 6--10 gold chunks; +4.55 pp on high-scatter Complex Reasoning, 71\%
win rate) and smallest for concentrated evidence (+1.85 pp). Heuristic
multi-hop traversal alone ($-$2.12 pp vs GraphRAG-local)
\emph{underperforms} single-hop expansion, confirming that the value
lies in the LLM's adaptive decisions --- not in multi-hop traversal per
se.

Cross-model validation reveals that the advantage is a function of
\textbf{tool-calling sophistication}. At matched scale (n=100), the
ordering is: Claude Haiku 4.5 (+3.60 pp, p \textless{} 0.001)
\textgreater{} Gemini 2.5 Flash Lite (+0.84 pp, p=0.52, NS)
\textgreater{} Gemma 4 E2B ($-$0.78 pp, p=0.45, NS). Claude is the only
controller to achieve a statistically significant advantage at this
scale. For context, Claude's full-scale result (n=519, +4.37 pp over
heuristic and +2.42 pp over GraphRAG, both p \textless{} 0.001) uses the
primary Gemini 2.0 Flash experiments as the comparison baseline; the
+2.47 pp Gemini number in the abstract and Section 5.5 is from n=519
with Gemini 2.0 Flash, not the n=100 cross-model run. These numbers come
from different experiments and should not be read as a single monotonic
ladder without that caveat. The Claude result is the paper's strongest:
it is the first controller to \textbf{statistically significantly
outperform GraphRAG-local}. Cross-scale results on MuSiQue show
cross-scale attenuation (\textasciitilde62\% smaller delta on 20-chunk
KGs) consistent with the theory. The Gemma 4 E2B result establishes both
a deployment boundary (net negative for current local models) and a
concrete distillation target: exploration traces from strong controllers
can serve as training signal for fine-tuning smaller models via TRL,
enabling on-device KG navigation without API dependencies.

The core architectural insight is the separation of \textbf{candidate
discovery} from \textbf{ranking}: let the LLM explore the mention graph
to assemble a broad candidate pool, then rank with pure vector
similarity. This resonates with recent work on adaptive computation
depth (Zhu et al., 2025), where looped transformer models allocate more
iterations to complex inputs. RLM-on-KG performs the analogous
optimization over external structure: allocating more tool-calling
iterations to questions whose evidence scatter demands deeper
exploration.

The approach is deployable on existing RDF knowledge graphs with
entity-centric identifiers and provenance links, without additional
offline indexing pipelines.

\textbf{In sum: LLM control is worth paying for when evidence is
structurally scattered and the controller can use tools well; otherwise,
cheaper graph retrieval remains sufficient.}

Code and data are available at https://github.com/wordlift/rlm-on-kg.

\begin{center}\rule{0.5\linewidth}{0.5pt}\end{center}

\subsection{References}\label{references}

\begin{itemize}
\item
  Buehler, M. J. (2024). Accelerating Scientific Discovery with
  Generative Knowledge Extraction, Graph-Based Representation, and
  Multimodal Intelligent Graph Reasoning. arXiv:2403.11996.
\item
  Carbonell, J. and Goldberg, J. (1998). The Use of MMR, Diversity-Based
  Reranking for Reordering Documents and Producing Summaries. SIGIR
  1998.
\item
  Edge, D. et al.~(2024). From Local to Global: A Graph RAG Approach to
  Query-Focused Summarization. arXiv:2404.16130.
\item
  GraphWalks (2025). GraphWalks: A Benchmark for Long-Context Graph
  Reasoning. OpenAI, github.com/openai/graphwalks. Released alongside
  GPT-4.1, April 2025.
\item
  Guo, Z. et al.~(2025). LightRAG: Simple and Fast Retrieval-Augmented
  Generation. Findings of EMNLP 2025, pp.~10746--10761.
\item
  Gutierrez, B. J. et al.~(2024). HippoRAG: Neurobiologically Inspired
  Long-Term Memory for Large Language Models. NeurIPS 2024.
\item
  Izacard, G. and Grave, E. (2021). Leveraging Passage Retrieval with
  Generative Models for Open Domain Question Answering. EACL 2021.
\item
  Kontonis, V. et al.~(2026). Memento: Teaching LLMs to Manage Their Own
  Context. Microsoft Research.
\item
  Lewis, P. et al.~(2020). Retrieval-Augmented Generation for
  Knowledge-Intensive NLP Tasks. NeurIPS 2020.
\item
  Ma, X. et al.~(2023). Query Rewriting for Retrieval-Augmented Large
  Language Models. EMNLP 2023.
\item
  Trivedi, H. et al.~(2022). MuSiQue: Multihop Questions via Single-hop
  Question Composition. TACL 2022.
\item
  Trivedi, H. et al.~(2023). Interleaving Retrieval with
  Chain-of-Thought Reasoning for Knowledge-Intensive Multi-Step
  Questions. ACL 2023.
\item
  Xiang, Y. et al.~(2025). When to use Graphs in RAG: A Comprehensive
  Analysis for Graph Retrieval-Augmented Generation. arXiv:2506.05690.
  Accepted at ICLR 2026.
\item
  Zhang, S., Kraska, T., and Khattab, O. (2025). Recursive Language
  Models. arXiv:2512.24601.
\item
  Zhu, R.-J. et al.~(2025). Scaling Latent Reasoning via Looped Language
  Models (Ouro). arXiv:2510.25741.
\end{itemize}

\begin{center}\rule{0.5\linewidth}{0.5pt}\end{center}

\subsection{Acknowledgments}\label{acknowledgments}

The authors used Antigravity to assist with the research and the writing
to improve clarity and readability. All content was carefully reviewed
and approved by the authors, who retain full responsibility for the
accuracy of the work and for any errors or omissions.

\end{document}